\documentclass[manuscript]{aastex}
\input{psfig}
\usepackage{graphicx}
\usepackage{amsmath}
\usepackage{amssymb}

\input blackdvi.tex

\begin{document}

\title{\large Origin of the Structure of the Kuiper Belt during a
  Dynamical Instability in the Orbits of Uranus and Neptune}

\author {\textbf{Harold F. Levison}}
\affil{\small\em Southwest Research Institute, USA}

\author{\textbf{Alessandro Morbidelli}} \affil{\small\em Observatoire
  de la C\^ote d'Azur, France}

\author{\textbf{Christa Van Laerhoven} \affil{\small\em Department
      of Physics and Astronomy, University of British Colombia}}

\author{\textbf{Rodney Gomes}}
\affil{\small\em Observat\'orio Nacional/MCT, Brazil}

\and

\author{\textbf{Kleomenis Tsiganis}} \affil{\small\em Department of
  Physics, Aristotle University of Thessaloniki, Greece}

\begin{abstract}
  
  We explore the origin and orbital evolution of the Kuiper belt in
  the framework of a recent model of the dynamical evolution of the
  giant planets, sometimes known as the {\it Nice} model. This model
  is characterized by a short, but violent, instability phase, during
  which the planets were on large eccentricity orbits.  It
  successfully explains, for the first time, the current orbital
  architecture of the giant planets (Tsiganis et al$.$~2005), the
  existence of the Trojans populations of Jupiter and Neptune
  (Morbidelli et al$.$~2005), and the origin of the late heavy
  bombardment of the terrestrial planets (Gomes et al$.$~2005).  One
  characteristic of this model is that the proto-planetary disk must
  have been truncated at roughly 30 to $35\,$AU so that Neptune would
  stop migrating at its currently observed location.  As a result, the
  Kuiper belt would have initially been empty.
  
  In this paper we present a new dynamical mechanism which can deliver
  objects from the region interior to $\sim\!35\,$AU to the Kuiper
  belt without excessive inclination excitation.  In particular, we
  show that during the phase when Neptune's eccentricity is large, the
  region interior to its 1:2 mean motion resonance becomes unstable
  and disk particles can diffuse into this area.  In addition, we
  perform numerical simulations where the planets are forced to evolve
  using fictitious analytic forces, in a way consistent with the
  direct N-body simulations of the Nice model.  Assuming that the last
  encounter with Uranus delivered Neptune onto a low-inclination orbit
  with a semi-major axis of $\sim\!27\,$AU and an eccentricity of
  $\sim\!0.3$, and that subsequently Neptune's eccentricity damped in
  $\sim\!1$~My, our simulations reproduce the main observed properties
  of the Kuiper belt at an unprecedented level.  In particular, our
  results explain, at least qualitatively: 1) the so-existence of
  resonant and non-resonant populations, 2) the
  eccentricity--inclination distribution of the Plutinos, 3) the
  peculiar semi-major axis -- eccentricity distribution in the
  classical belt, 4) the outer edge at the 1:2 mean motion resonance
  with Neptune, 5) the bi-modal inclination distribution of the
  classical population, 6) the correlations between inclination and
  physical properties in the classical Kuiper belt, the existence of
  the so-called extended scattered disk.  Nevertheless, we observe in
  the simulations a deficit of nearly-circular objects in the
  classical Kuiper belt.
\end{abstract}

\section{Introduction}

The Kuiper belt is the relic of the primordial planetesimal disk,
shaped by various dynamical and collisional processes that occurred
when the Solar System was evolving towards its present structure.
Thus, studying the origin of the structure of the Kuiper belt is
important because it can unveil the history of the formation and
evolution of the giant planets and, more in general, of the
proto-Solar System.

\begin{figure}[t!]
\centerline{\psfig{figure=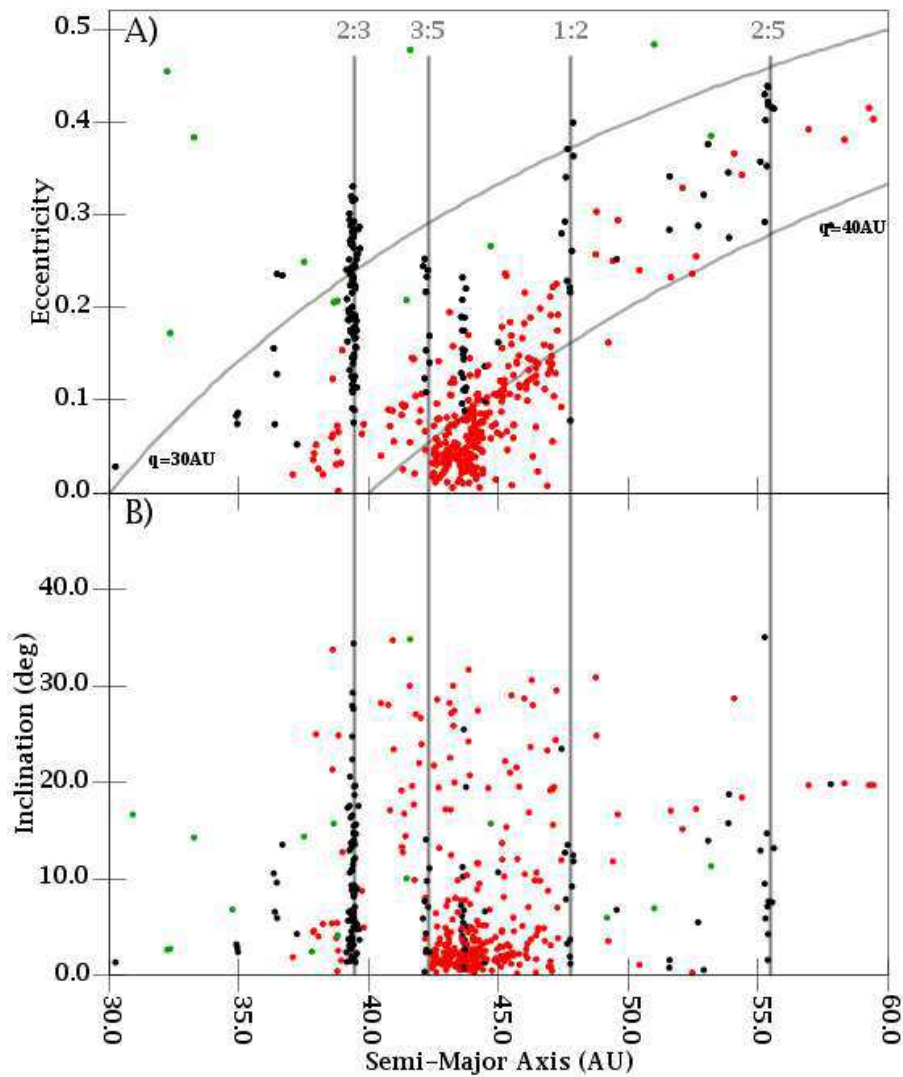,height=14.4cm}}
\vspace*{-.3cm}
\caption{Observed orbital elements distribution  of
  the (KBOs) with orbits determined from observations over at least
  three oppositions. We have employed the methods in Gladman et
  al$.$~(2007) to classify these objects according to their dynamical
  behavior.  In particular, the dots that are black represent objects
  in one of Neptune's mean motion resonances, dots that are green are
  objects undergoing encounters with Neptune, and the red dots are
  non-resonant stable objects. A) Eccentricity versus semi-major axis.
  The two curves correspond to $q=30$~AU and $q=40$.  B) semi-major
  axis versus inclination.}  \vspace{0.3cm}
\label{obs}
\end{figure}

The main properties of the Kuiper belt that require an explanation in
the framework of the primordial evolution of the Solar System are:
(The following list is presented in no particular order.)
\begin{itemize}
  
\item[i)] The existence of conspicuous populations of objects in the
  main mean motion resonances (MMRs) with Neptune (2:3, 3:5, 4:7, 1:2,
  2:5, etc.). Resonant objects form obvious vertical structures in a
  semi-major axis ($a$) versus eccentricity ($e$) plot, for example,
  see Fig.~\ref{obs}A.  The resonant objects represent a significant
  fraction of the total trans-neptunian population.  Trujillo et
  al$.$~(2001) estimate that roughly 10\% of the total population of
  the Kuiper belt are in the resonances, while Kavelaars et
  al$.$~(2007) put the fraction of objects in Neptune's 2:3~MMR alone
  at $\sim\!20\%$.
  
\item[ii)] The excitation of the eccentricities in the classical belt,
  which we define as the collection of stable, non-resonant objects
  with $a\!<\!48$~AU.  The median eccentricity of the classical belt
  is $\sim\!0.07$.  This value is small, but nevertheless it is an
  order of magnitude larger than what must have existed when the
  Kuiper belt objects (KBOs) formed (Stern \& Colwell~1997a; Kenyon \&
  Luu~1998 1999a; 1999b).  It should be noted, however, that the upper
  eccentricity boundary of this population (see Fig.~\ref{obs}A) is
  set by the long-term orbital stability in this region (Duncan et
  al$.$~1995), and thus the classical belt could have originally
  contained objects with much larger eccentricities.
  
\item[iii)] The $a$--$e$ distribution of classical belt objects (see
  Fig.~\ref{obs}A) shows another distinct feature that models must
  explain.  The population of objects on nearly-circular orbits
  effectively `ends' at about 44~AU, and beyond this location the
  eccentricity tends to increase with semi-major axis (see Kavelaars
  et al$.$~2007 for a discussion). The lower bound of the bulk of the
  $a$--$e$ distribution in the 44--48 AU range follows a curve that is
  steeper than a curve of constant $q$.  Given that the observational
  biases are roughly a function of $q$, the apparent relative
  under-density of objects at low eccentricity in the region
  immediately interior to Neptune's 1:2 MMR is likely to be a real
  feature of the Kuiper belt distribution.
  
\item[iv)] The outer edge of the classical belt (Fig.~\ref{obs}A).
  This edge appears to be precisely at the location of the 1:2~MMR
  with Neptune.  Again, the under density (or absence) of low
  eccentricity objects beyond the 1:2 MMR cannot be explained by
  observational biases (Trujillo \& Brown~2001; Allen et al$.$~2001,
  2002).
  
\item[v)] The inclination distribution in the classical belt.
  Fig.~\ref{obs}B shows a cluster of objects with $i\lesssim 4^\circ$,
  but also several objects with much larger inclinations, up to $i\sim
  30^\circ$. Observational biases definitely enhance the low
  inclination cluster relative to the large inclination population
  (the probability of discovery of an object in an ecliptic survey is
  roughly proportional to $1/\sin(i)$). However, the cluster persists
  even when the biases are taken into account.  Brown~(2001) argued
  that the de-biased inclination distribution is bi-modal and can be
  fitted with two Gaussian functions, one with a standard deviation
  $\sigma \sim 2^\circ$ for the low-inclination core, and the other
  with $\sigma \sim 12^\circ$ for the high inclination population.
  Since the work of Brown, the classical population with $i<4^\circ$
  is called `cold population', and the higher inclination classical
  population is called `hot population'.  It is interesting to note
  that the fact that most known classical belt objects are members of
  the cold population is a result of observational biases.  The hot
  population actually dominates in total number.
  
\item[vi)] The correlations between physical properties and orbital
  distribution. The cluster of low inclination objects visible in the
  ($a,i$) distribution disappears if one selects only objects with
  absolute magnitude $H\lesssim6$ (Levison \& Stern~2001).  This
  implies that intrinsically bright objects are under represented in
  the cold population.  Grundy et al$.$~(2005) have shown that the
  objects of the cold population have a larger albedo, on average,
  than those of the hot population. Thus, the correlation found by
  Levison and Stern implies that the hot population contains bigger
  objects.  Bernstein et al$.$~(2004) showed that the hot population
  has a shallower $H$ distribution than the cold population, which is
  consistent with the absence of the largest objects in the cold belt.
  
\item[] In addition, there is a well known correlation between color
  and inclination (Tegler \& Romanishin~2000; Doressoundiram et
  al$.$~2001; Trujllo \& Brown~2002; Doressoundiram et al$.$~2005;
  Elliot et al$.$~2005).  The hot population objects show a wide range
  of colors, from red to gray.  Conversely, the cold population
  objects are mostly red. In other words, the cold population shows a
  significant deficit of gray color bodies relative to the hot
  population. The same is true for the objects with $q\!>\!39$~AU
  (Doressoundiram et al$.$~2005). These objects are all red, regardless
  of their inclination. If one includes also objects with smaller $q$,
  then a correlation between color and perihelion distance becomes
  apparent (gray color objects becoming more abundant at low $q$).
  
\item[vii)] The existence of the {\it extended scattered disk}, which
  consists of stable non-resonant objects with semi-major axes beyond
  Neptune's 1:2MMR and mainly have perihelion distances between
  $\sim\!30$ and $\sim\!40\,$AU.  These objects have also been called
  {\it detached objects.}  They cannot have been placed on their
  current orbits by the current configuration of the planets and thus
  supply an important constraint for formation models. It is important
  to note that we do not consider Sedna to be a member of the extended
  scattered disk.  We believe that this object represents a totally
  different population that had a different origin (Mordibelli \&
  Levison~2004; Kenyon \& Bromley~2004a; Brasser et al.~2007).

\item[viii)]The mass deficit of the Kuiper belt. The current mass of the
  Kuiper belt is very small --- estimates range from 0.01~$M_\oplus$
  (Bernstein et al$.$~2004) to 0.1~$M_\oplus$ (Gladman et al$.$~2001).
  The uncertainty is due mainly to the conversion from absolute
  magnitudes to sizes, assumptions about bulk density, and ambiguities
  in the size distribution.  Whatever the exact mass really is, there
  appears to be a significant mass deficit (of 2--3 orders of
  magnitude) with respect to what models say was needed in order for
  the KBOs to accrete where we see them. In particular, the growth of
  these objects within a reasonable time ($10^7$--$10^8$~My) requires
  the existence of about 10 to 30~$M_\oplus$ of solid material in a
  dynamically cold disk (Stern, 1996; Stern \& Colwell, 1997a, 1997b;
  Kenyon \& Luu, 1998; 1999a; 1999b; Kenyon \& Bromley~2004b).
\end{itemize}

A good model of the Kuiper belt's primordial evolution should explain
all the properties detailed above. In the simplest scenario, the main
features of the belt are the consequence of the outward migration of
Neptune, due to its interaction with the primordial planetesimal disk.
In this scenario, the planetesimal disk, which was initially
dynamically cold, extended to at least $\sim 50$~AU.  Neptune migrated
on a nearly-circular, low-inclination orbit, from an unconstrained
initial location (estimates ranging from 18~AU --- Gomes~2003 --- to
23~AU --- Malhotra~1995) up to its current orbital radius of 30~AU.

According to the above scenario, three main events happen in the
Kuiper belt during this time. The main mean motion resonances with the
Neptune, which moved along with the planet, captured objects from the
cold distant disk that they swept through (Malhotra~1993, 1995; Hahn
\& Malhotra 1999, 2005).  Simultaneously, a large fraction of the
objects initially in the region swept by Neptune's motion were
scattered by the planet onto orbits with large eccentricity and
semi-major axis. The relic of this population is now called the {\it
  scattered disk}. Finally, a small fraction of the scattered disk
objects decoupled from the planet, decreasing their eccentricities
through interactions with some secular or mean-motion resonances
(Gomes~2003). If Neptune were not migrating, the decoupled objects
would soon, once again, have evolved back onto Neptune-crossing
orbits, because the dynamics are time reversible.  However, Neptune's
migration broke the reversibility, and some of the decoupled bodies
could manage to escape from the resonances and remained permanently
trapped in the Kuiper belt.  Gomes showed that the majority of these
trapped bodies have large inclinations (acquired during the scattering
phase), and identified them with the current hot Kuiper belt
population.  Some scattered disk bodies could also be trapped in the
main MMR with Neptune (Gomes~2003), and mixed with those captured from
the cold disk to form the current resonant populations.

Thus, in this view both the resonant populations and the hot
population are the product of Neptune's migration, whereas the cold
population is the local, original Kuiper belt population which was
only marginally perturbed during the resonance sweeping. Assuming that
the physical properties of the bodies (colors and maximal sizes)
varied in the primordial disk with heliocentric distance, this
scenario qualitatively explains why the scattered objects and hot
classical belt objects --- which mostly come from regions inside
$\sim\!30\,$AU --- appear to have similar color and size distributions,
while the cold classical objects --- the only ones that actually
formed in the trans-Neptunian region --- have different distributions.

Nevertheless, this scenario does not explain why the cold Kuiper belt
did not retain its primordial mass, nor why its outer edge coincides
with the location of Neptune's 1:2 MMR. To solve these problems,
Levison \& Morbidelli~(2003) proposed that the original planetesimal
disk, in which the planets and KBOs formed, was truncated near
$30\,$AU, and {\it all} the KBOs that we see were implanted in the
Kuiper belt during Neptune's migration. This idea also has the
advantage of explaining why Neptune stopped migrating at $30\,$AU ---
it hit the edge of the disk (Gomes et al$.$~2004).  In Levison \&
Morbidelli's scenario, the Kuiper belt's cold population was pushed
outward from interior to $30\,$AU with the following mechanism.  When
Neptune was close enough to the Sun for its 1:2 MMR to be within the
planetesimal disk, disk particles were trapped in this resonance as
Neptune migrated (Malhotra~1995).  As the resonance left the disk, it
pushed the particles outward along with it.  As these objects were
driven from the Sun many had small eccentricities because of a newly
discovered secular interaction with Neptune and they were
progressively released from the resonance due to the non-smoothness of
the planet's migration.  The 1:2 MMR does not excite inclinations, so
that these bodies could retain their initial low inclinations during
this process (hence the cold belt is cold). Another strength of this
scenario is that the final edge of the cold belt naturally coincides
with the final (and thus current) position of the 1:2 MMR.  The low
mass of the cold population is explained by the low efficiency of the
transport/implantation process.

In the light of Gomes~(2003) and Levison \& Morbidelli~(2003), the
most important properties of the Kuiper belt seem to be explained, at
least qualitatively.  Therefore, is there any need to re-investigate
the problem? We think that the answer is yes, for at least three
reasons.  First, the idea that both the cold and the hot populations
formed closer to the Sun and were transported outward re-opens the
issue of the correlation between physical properties and final orbital
distribution. Second, a quantitative comparison between the observed
orbital distribution of the Kuiper belt and that expected from the
sculpting model (including observational biases) has never been done.
Third, and most importantly, our view of the evolution of the giant
planets' orbits has changed radically.

Last year, we proposed a new comprehensive scenario --- now often
called `the Nice model' because we were all working in the city of
Nice when the model was developed --- that reproduces, for the first
time, many of the characteristics of the outer Solar System.  It
quantitatively recreates the orbital architecture of the giant planet
system (orbital separations, eccentricities, inclinations; Tsiganis et
al$.$~2005) and the capture of the Trojan populations of Jupiter
(Morbidelli et al$.$~2005) and Neptune (Tsiganis et al$.$~2005;
Sheppard \& Trujillo~2006) and many of the irregular satellites
(Nesvorn\'y et al$.$~2006). It also naturally supplies a trigger for
the so-called Late Heavy Bombardment (LHB) of the terrestrial planets
that occurred $\sim\!3.8$ billion years ago \Red{(Tera et
  al$.$~1974)}, and quantitatively reproduces most of the LHB's
characteristics \Red{(Gomes et al$.$~2005)}. The evolution of the
giant planets in the Nice model is substantially different from the
smooth migration on low eccentricity, low inclination orbits,
envisioned in Malhotra~(1995), Gomes~(2003), and Levison \&
Morbidelli~(2003).  Consequently, it is important to re-examine, from
scratch, the issue of the primordial shaping of the Kuiper belt in the
framework of the Nice model. This study could bring additional
support, or refutation, of the Nice model, thus deepening our
understanding of the processes at work in the early Solar System.

This paper is structured as follows. In section~\ref{nice}, we review
the Nice model, stressing the aspects that may be important for the
structure of the Kuiper belt. In section~\ref{newpush} we discuss a
new mechanism that could be effective for the transport of objects
from the primordial planetesimal disk into the Kuiper belt region. The
remaining part of the paper presents the results of our simulations.
Section~\ref{dist} focuses on the resulting $a$--$e$ and
$i$-distributions. Section \ref{correlations} addresses the origin of
the correlations between the physical and the orbital properties.
Section~\ref{discussion} discusses several important issues such as:
the efficiency of the overall transport mechanism and the origin of
the mass deficit of the Kuiper belt (sect.~\ref{mass}); the orbital
distribution of the Plutinos (sect.~\ref{2to3}); and the filling of
mean motion resonances beyond 50~AU (sect.~\ref{dist-res}).
Section~\ref{conclusions} summarizes the main results, stressing
advantages and weak points of our model.

\section{The `Nice' model of evolution of the giant planets}  
\label{nice}

The initial conditions of the Nice model are intended to represent the
state of the outer Solar System at the time of disappearance of the
gas disk.  The giant planets are assumed to be initially on
nearly-circular and coplanar orbits, consistent with the expectations
from the theory of formation of giant planets (Pollack et al.~1996;
Lubow et al.~1999). A pre-migration configuration is assumed, with
planetary orbital separations that are significantly smaller than
those currently observed.  More precisely, the giant planet system is
assumed to be in the range from $\sim 5.5$~AU to $\sim 14$~AU. The gas
giants (Jupiter and Saturn) are placed closer to the Sun than the ice
giants (Uranus and Neptune).  Saturn is assumed to be closer to
Jupiter than their mutual 1:2 MMR, a condition required in order to
avoid a substantial amount of migration (Type~II) during the gas disk
lifetime (Masset \& Snellgrove~2001; Morbidelli \& Crida~2007).
Overall, such a compact system is consistent with the constraints on
the formation timescales of Uranus and Neptune (Levison \&
Stewart~2001; Thommes et al.~2003).

A planetesimal disk is assumed to exist beyond the orbits of the giant
planets.  In particular, it was assumed that particles inhabited only
those regions where the dynamical lifetime of the individual objects
is of the order of the gas disk lifetime ($\sim 3$~My; see for
example, Haisch, Lada \& Lada 2001), or longer, because the
planetesimals initially on orbits with shorter dynamical lifetimes
should have been eliminated during the nebula era.  This sets the
inner edge of the disk to be about 1.5~AU beyond the location of the
outermost planet.  The outer edge of the disk is assumed at $\sim
34$~AU.  It was found that in order to most accurately reproduce the
characteristics of the outer planetary system the total mass of the
disk must have been $\sim\!35 M_\oplus$.

\begin{figure}[t!]
\centerline{\psfig{figure=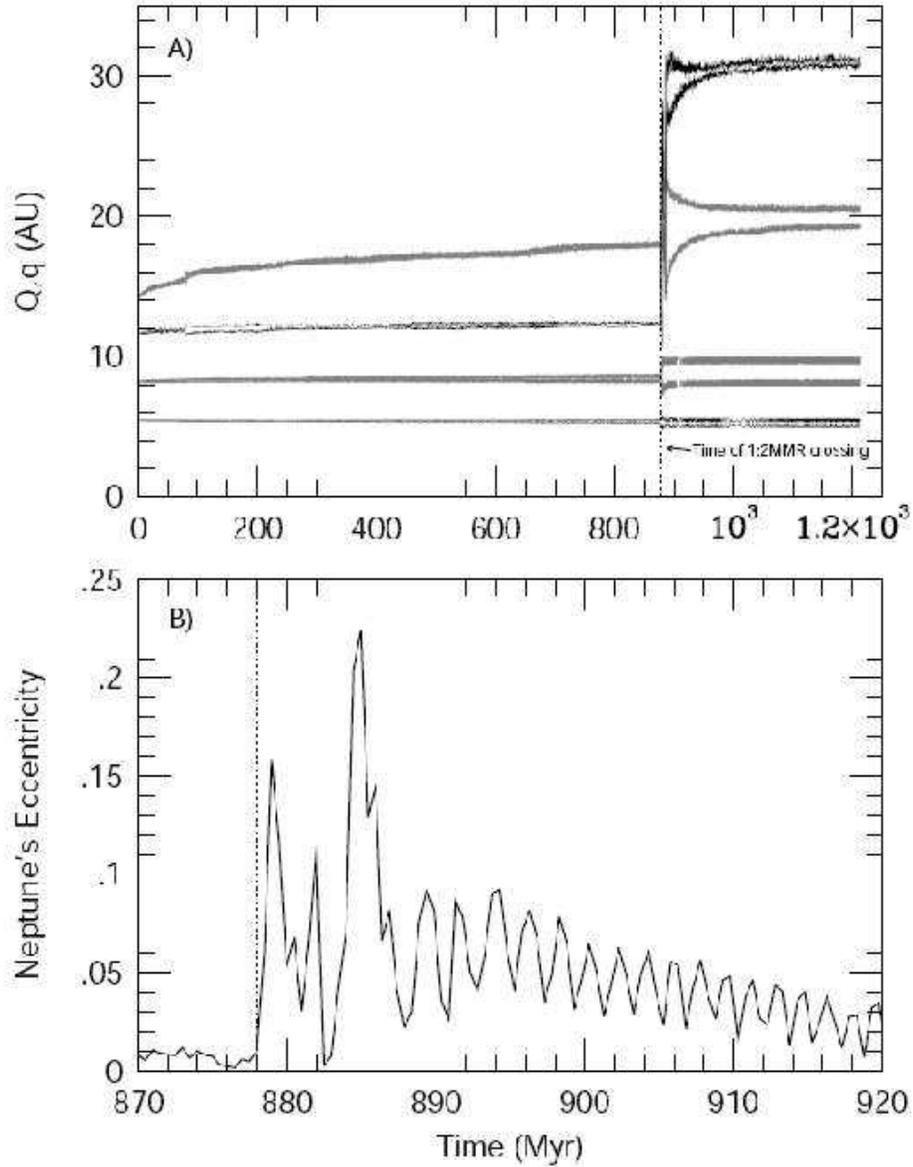,height=16.8cm}}
\vspace*{-.3cm}
\caption{Giant planet evolution in the Nice model. A) Each planet is
  represented by a pair of curves -- the top and bottom curves are the
  aphelion and perihelion distances, respectively.  In this simulation
  Jupiter and Saturn cross their 1:2 mean-motion resonance at 878~My,
  which is indicated by the dotted line (From Gomes et al$.$~2005). B)
  The eccentricity of Neptune.}

\vspace{0.3cm}
\label{LHB}
\end{figure}

With the above configuration, the planetesimals in the inner regions
of the disk acquire planet-scattering orbits on a timescale of a few
million years.  Consequently, the migration of the giant planets
proceeds at very slow rate, governed by the slow planetesimal escape
rate from the disk (Fig.~\ref{LHB}A).  After a significant period of
time, ranging from 60 My to 1.1~Gy in the simulations in Gomes et
al$.$~(2005), Jupiter and Saturn eventually cross their mutual 1:2
mean-motion resonance.  (The upper range of this time-span is
consistent with the timing of the LHB, which occurred approximately
650~My after planet formation.)  The resonance crossing excites their
eccentricities to values slightly larger than those currently
observed.

The small jump in Jupiter's and Saturn's eccentricities drives up the
eccentricities of Uranus and Neptune to the point where they start to
approach each other.  Thus, a short phase of violent encounters
follows the resonance-crossing event (from $\sim\!878$ to
$\sim\!885\,$Myr in the example shown in Fig.~\ref{LHB}).
Consequently, both ice giants are scattered outward, onto large
eccentricity orbits ($e\sim$~0.25--0.4) that penetrate deeply into the
disk. This destabilizes the full planetesimal disk and disk particles
are scattered all over the Solar System.  The eccentricities of Uranus
and Neptune and, to a lesser extent, of Jupiter and Saturn, are damped
on a timescale of a few My due to the dynamical friction exerted by
the planetesimals.  In the example shown in Fig.~\ref{LHB}, Neptune's
eccentricity evolves from a peak of 0.22 to less than 0.05 in
$2.4\,$Myr. Tsiganis et al$.$~(2005) found that this damping time is
between $\sim\!0.3$ and $\sim4\,$Myr.

As a result of the gravitational influence of the disk, the planets
decouple from each other, and the phase of mutual encounters rapidly
ends. During and after the eccentricity damping phase, the giant
planets continue their radial migration, and eventually reach final
orbits when most of the disk has been eliminated (Fig.~\ref{LHB}A).
The final outcomes of the simulations of the Nice model reproduce
quantitatively the current architecture of the giant planets, in terms
of semi-major axes, eccentricities, and inclinations (Tsiganis et
al$.$~2005).

The sudden destabilization of the planetesimal disk produces an abrupt
spike in the flux of bodies from the outer Solar System entering the
region of the terrestrial planets.  In addition, the rapid migration
of Jupiter and Saturn from their mutual 1:2 MMR to their current
position, destabilizes approximately 90\% of the asteroids residing in
the asteroid belt at the time. Together, outer Solar System
planetesimals and escaping asteroids cause a short-lived bombardment
on the terrestrial planets lasting $\lesssim\!100$~My --- the
magnitude of which is consistent with constraints on the lunar crater
rate at $\sim\!3.8\,$Ga (Gomes et al$.$~2005).

Moreover, a fraction of the disk planetesimals are trapped onto Jovian
Trojan orbits as Jupiter and Saturn migrate away from the 1:2 MMR. The
orbital distribution of the trapped Trojans and their total mass is
remarkably similar with the observed distribution and the total
estimated population (Morbidelli et al$.$~2005). Neptune's Trojans are
also captured during the giant planet evolution, over a broad range of
inclinations (Tsiganis et al$.$~2005), consistent with the inclination
distribution of the newly discovered objects (although these are only
4 known objects; Trujillo \& Sheppard~2006).  Finally, disk particles
can be captured into orbit around Saturn, Uranus, and Neptune during
planet-planet encounters, thereby creating at least some of the
observed irregular satellites (Nesvorn\'y et al$.$~2006).

The unprecedented success of the Nice model calls for a
re-investigation of the formation of the Kuiper belt.  An essential
ingredient of the model is that, at least at the time of the LHB, the
planetesimal disk was truncated at 30--35~AU, otherwise Neptune would
have migrated too far.  This is consistent with the idea, first
presented in Levison \& Morbidelli~(2003), that Kuiper belt objects
that we see were formed inside this boundary and that the full Kuiper
belt --- both cold and hot populations --- was pushed outwards during
the evolution of the planets.  Since this idea so nicely solves the
Kuiper belt's mass depletion problem, our goal is to determine whether
it is still viable in light of the Nice model.  The issue is that
Neptune's evolution, as envisioned by this model, is substantially
different from the smooth low-eccentricity migration contemplated in
previous works.

The mechanism in Gomes (2003) for the push out of the hot populations
might still work in the framework of the Nice model.  Indeed, a
massive scattered disk is produced and Neptune has a final phase of
slow migration on circular orbit, which are the essential ingredients
of Gomes' mechanism.  However, the mechanism proposed by Levison \&
Morbidelli~(2003) for pushing out the cold population is in
significant trouble. For this mechanism to work, Neptune's 1:2 MMR
must have been within the disk as Neptune began a phase of smooth
outward migration.  Since the outer edge of the disk is at
$\sim\!34\,$AU, this implies that the semi-major axis of Neptune had
to have been within $\sim 21$~AU at this time. The semi-major axis of
Neptune after the last encounter with Uranus varies greatly from
simulation to simulation of the Nice model (even if the final position
of the planet is systematically at $\sim 30$~AU).  However, in none of
the successful simulations that we have produced did Neptune ever have
a semi-major axis as small as 21~AU when it stopped having encounters
with the other planets.

Therefore, we need to find another mechanism for the implantation of
the cold population in the Kuiper belt. This mechanism is detailed in
the next section and makes use of the new aspect of Neptune's
evolution: a transient phase when its eccentricity was large.
 
\section{A new transport mechanism to fill the Kuiper belt}
\label{newpush}

In this section we describe a new mechanism for the outward transport
of material from the primordial proto-planetary disk interior to
$30\,$AU to the current Kuiper belt.  This new mechanism is based on a
well-known characteristic of the trans-Neptunian regions --- Neptune's
MMRs are sticky (Holman \& Wisdom~1993; Levison \& Duncan~1997).  In
particular, integrations of test particles in the current scattered
disk show that objects commonly become temporarily trapped in MMRs
with Neptune, which can drive the particle's eccentricity to low
values, thereby decoupling them from Neptune (for example, see Levison
\& Duncan~1997, Fig$.$~7). In the current Solar System, these
decoupling events can only occur at very specific and narrow ranges of
semi-major axes because the mean motion resonances are narrow.  In
addition, since the orbits of the planets are not evolving, the
process is time reversible and so there is no permanent capture.

\begin{figure}[t!]
\centerline{\psfig{figure=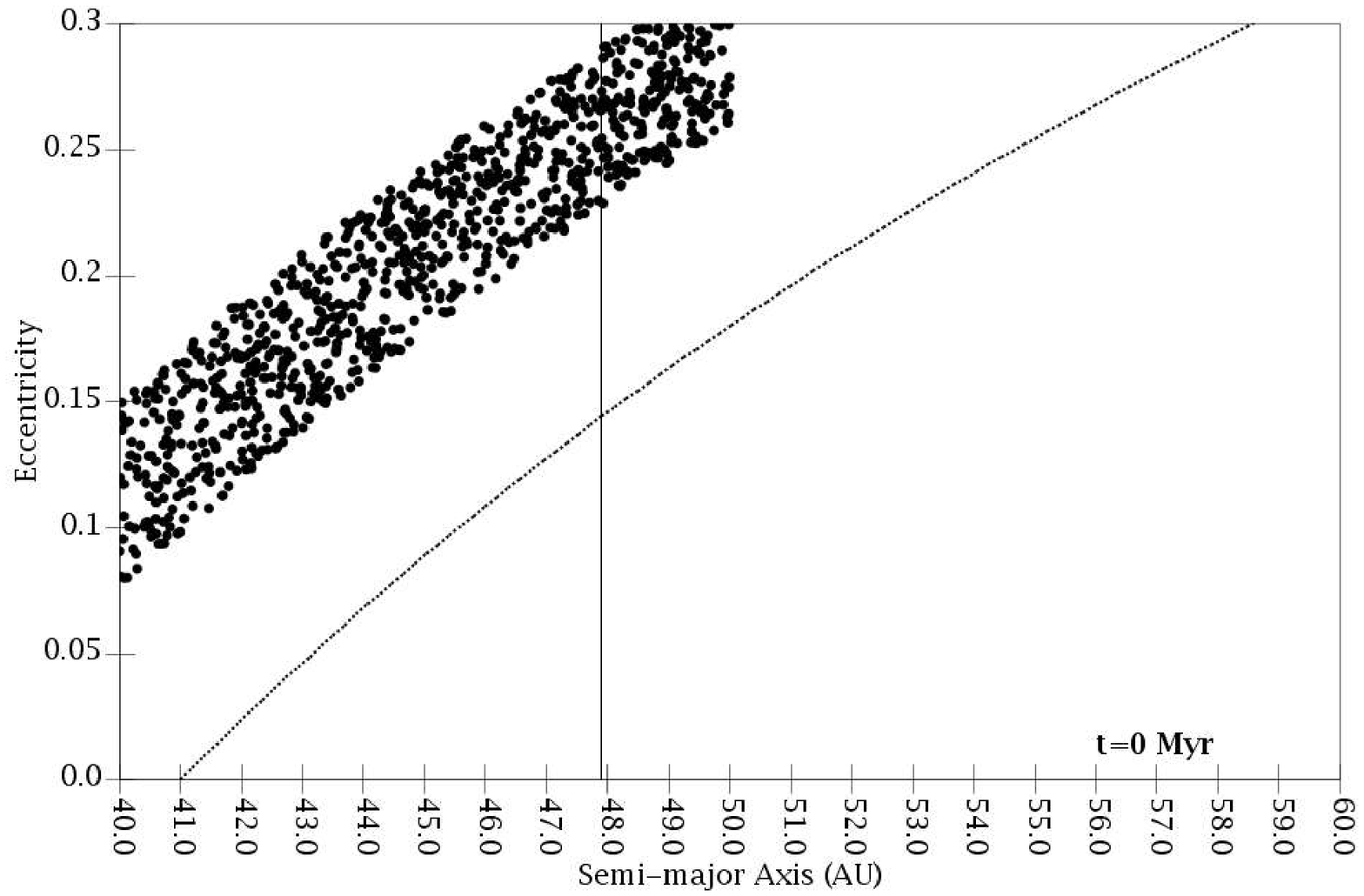,height=4.cm}
\psfig{figure=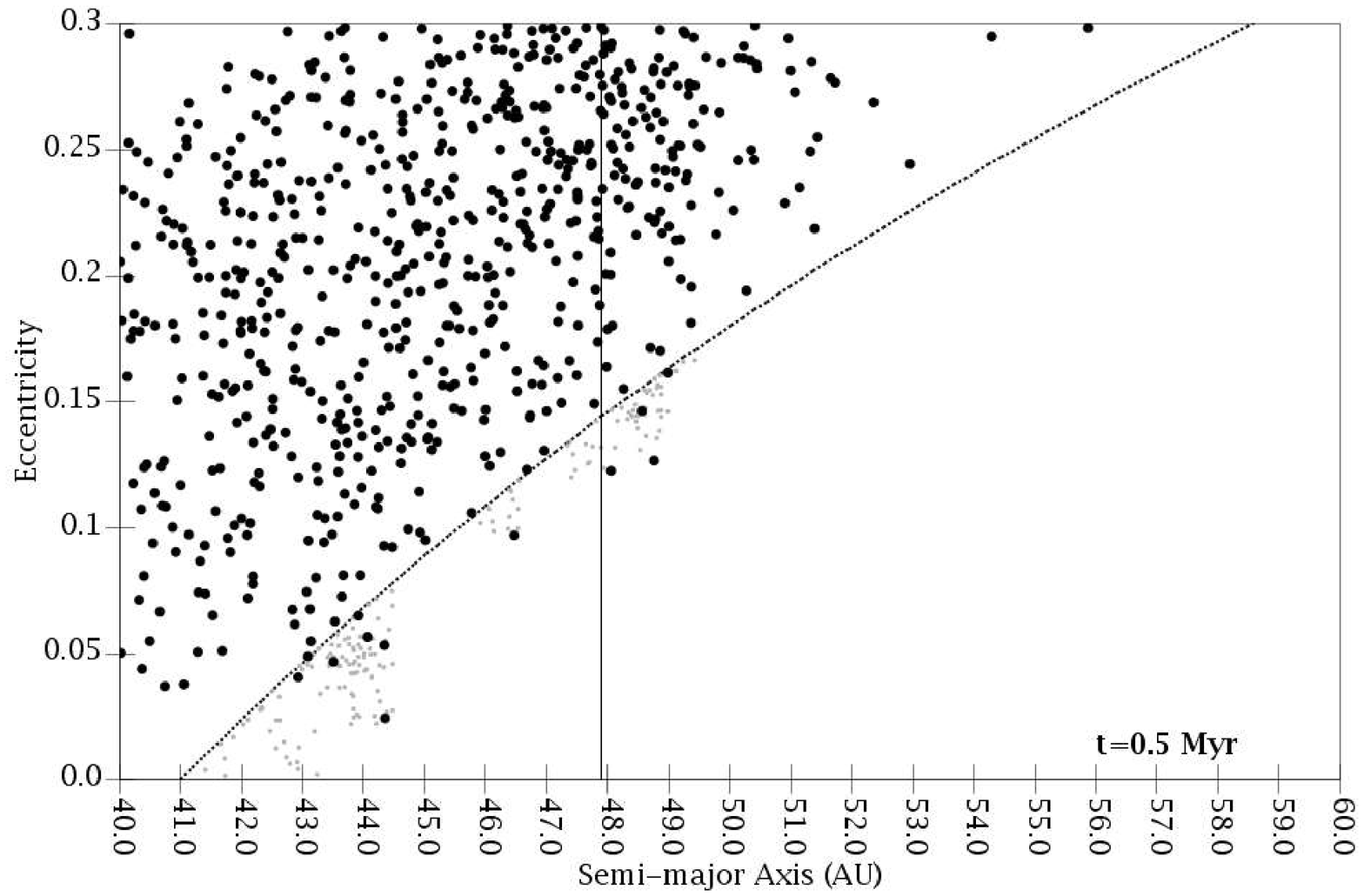,height=4.cm}}
\centerline{\psfig{figure=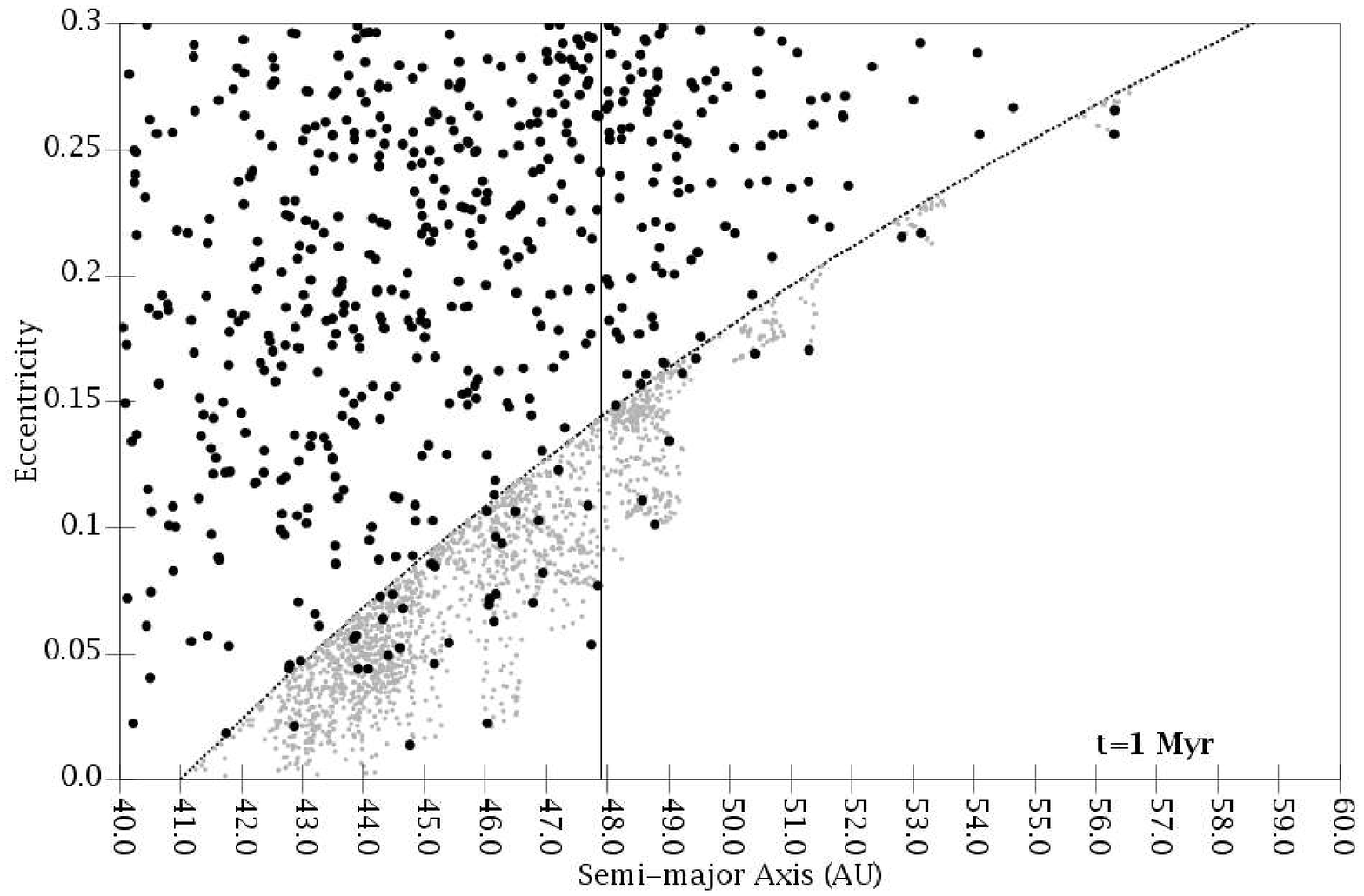,height=4.cm}
\psfig{figure=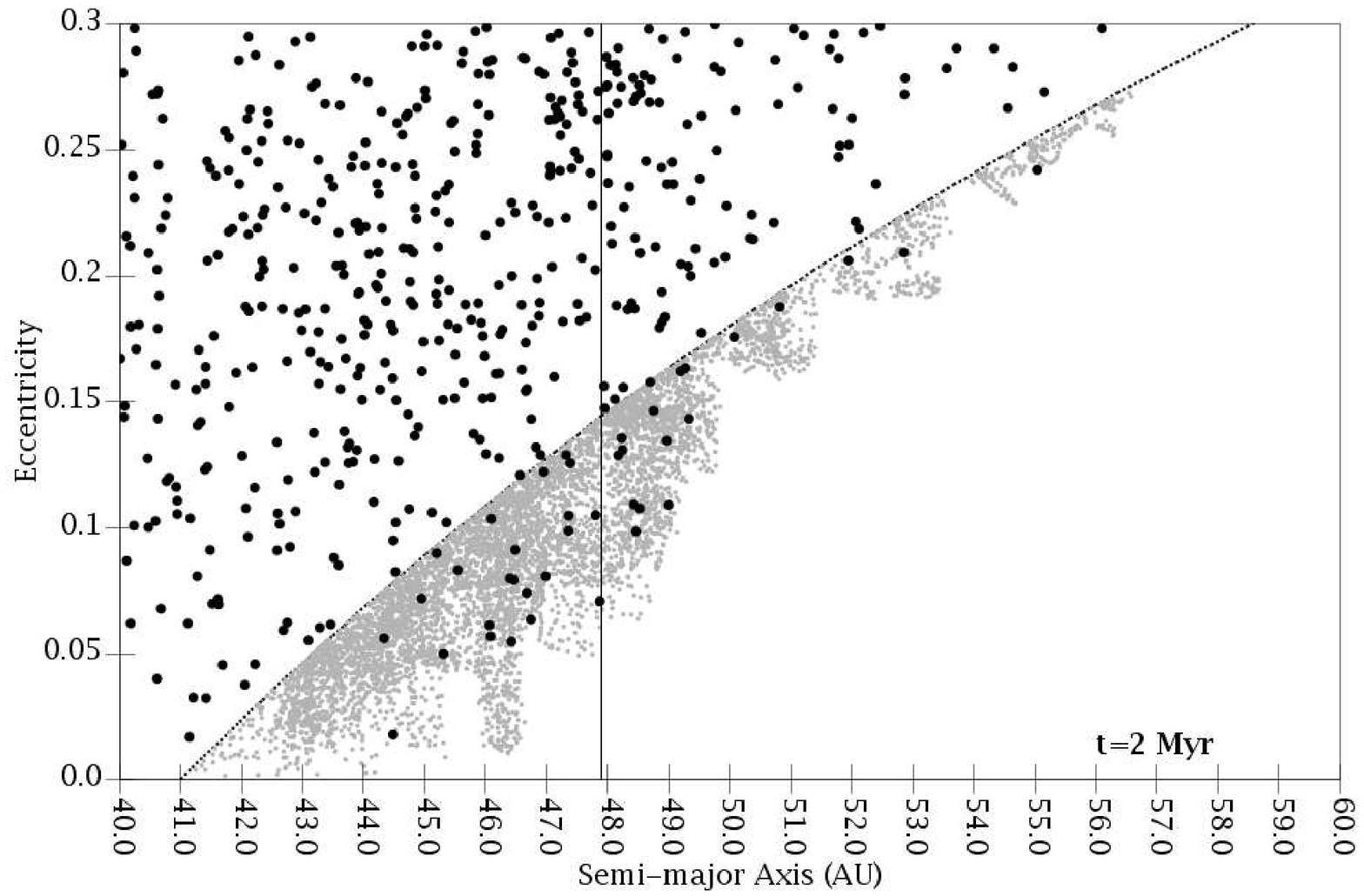,height=4.cm}}
\centerline{\psfig{figure=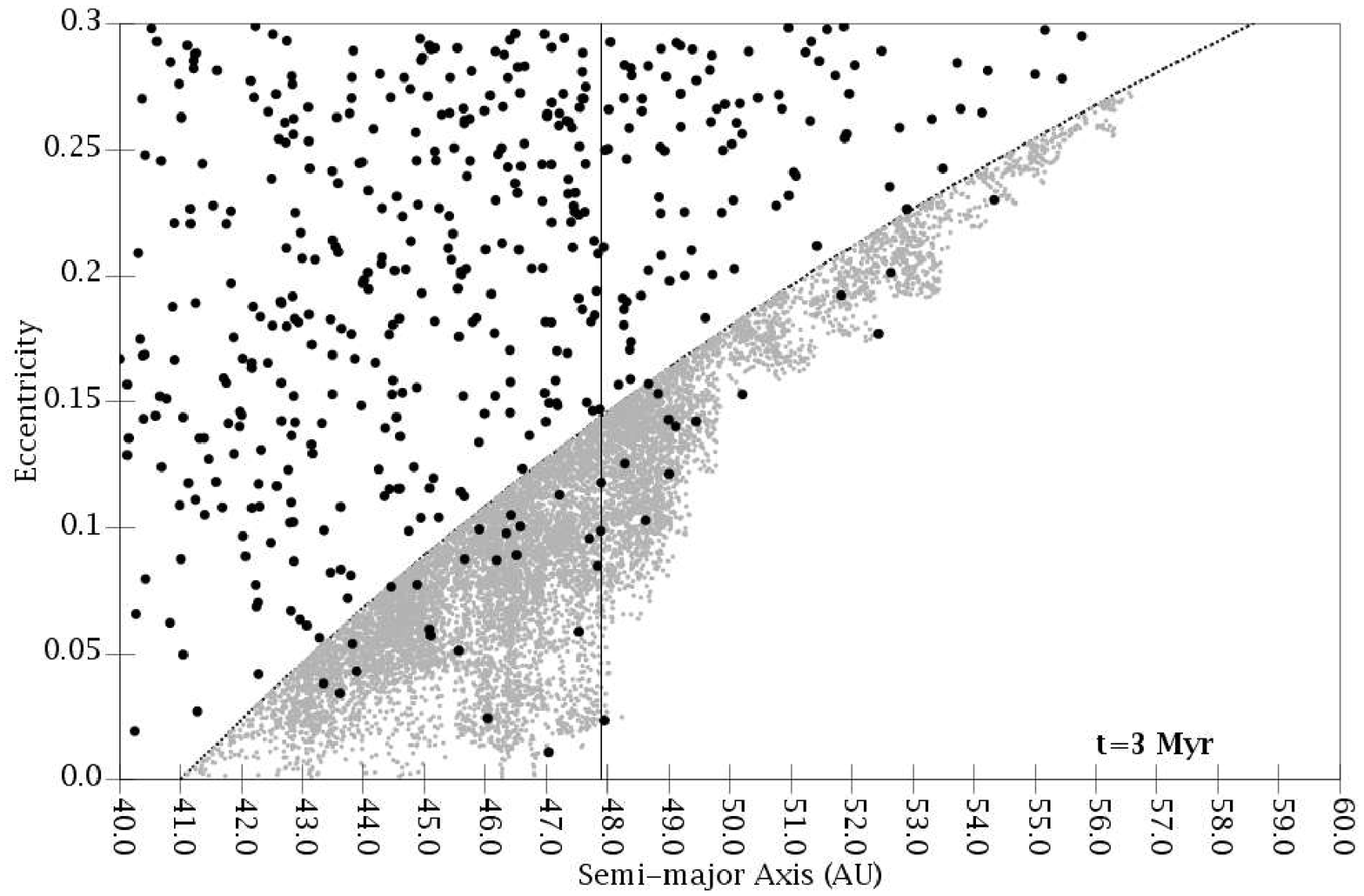,height=4.cm}
\psfig{figure=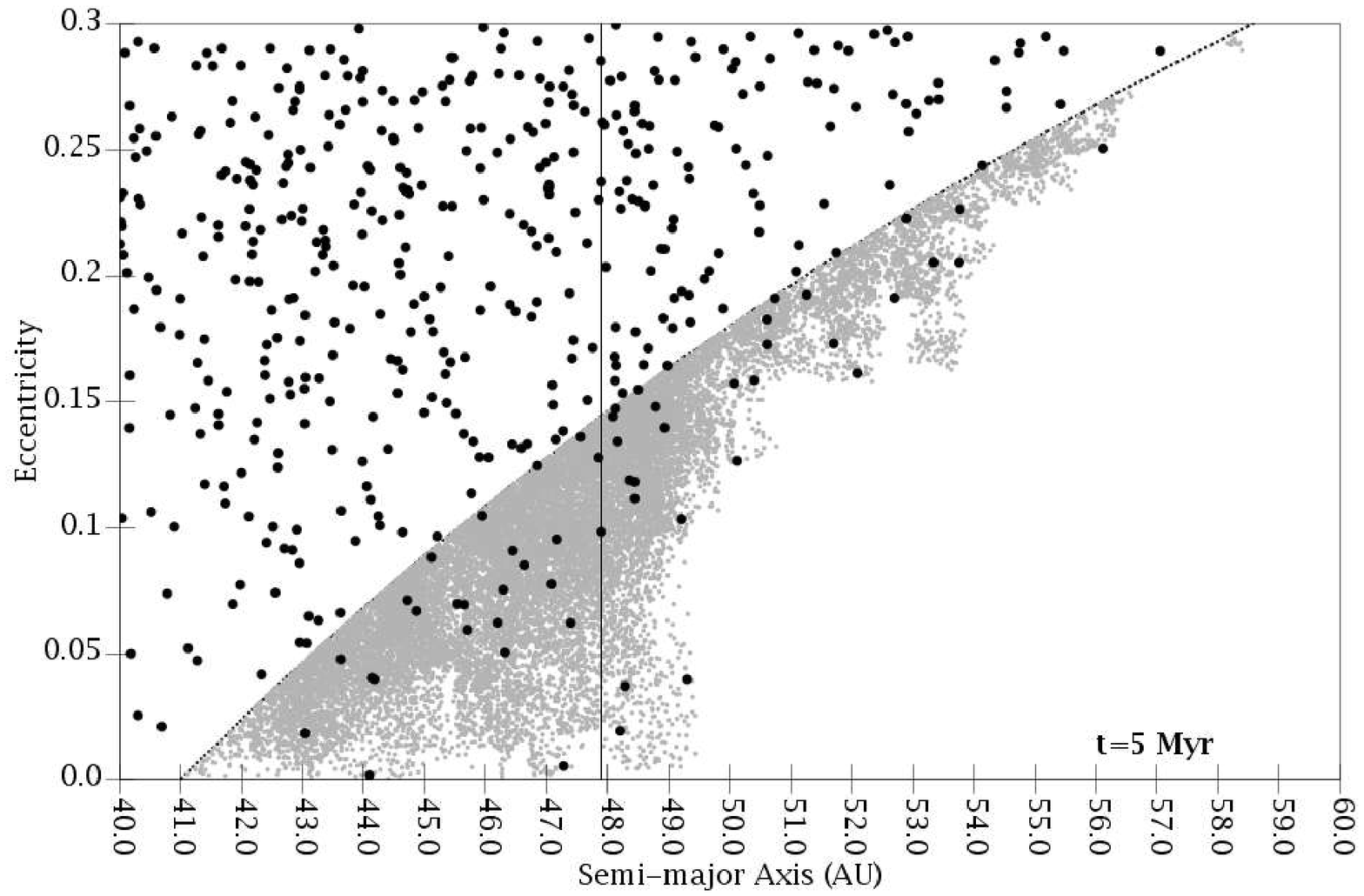,height=4.cm}}
\caption{\Black{Evolution of particles undergoing
    perturbations from Neptune on an eccentric orbit ($a=30$~AU,
    $e=0.2$). The big dots represent the test particles, initially all
    on Neptune crossing orbits. The solid curve marks $q=42\,$AU and
    the dotted vertical line the location of the 1:2 MMR with Neptune.
    The area cumulatively visited by the particles in the $q>42$~AU
    region is colored with small gray dots. Time evolves from the top
    left to the lower right panel.  Because of overlapping resonances,
    particles can evolve into the Kuiper belt and acquire orbits with
    $e\!\sim\!0$.  In addition, the 1:2 MMR is a natural boundary of
    the visited region.}}
\label{SD-invade}
\end{figure}

However, the situation looks very different if Neptune's orbit is
eccentric because the region of semi-major over which objects can be
decoupled from Neptune is much larger.  This occurs for two reasons.
First, Neptune's large eccentricity forces extensive secular
oscillations in the eccentricity of a particle in this region, thereby
allowing it's orbit to become temporally nearly circular at some
locations.  We find that for the values of eccentricity that we see in
the Nice model simulations, these secular effects can be important in
the 40 to $50\,$AU region.  However, it takes longer than $2\,$Myr for
the eccentricity to drop and thus this process is probably only
important when Neptune's damping time is long.

More importantly, when Neptune's eccentricity is large the widths of
all its mean motion resonances increase (see Morbidelli~2002).
Indeed, numerical experiments of the scattering process show that, for
eccentricities larger than $\sim\!0.15$, the MMRs interior to the
1:2MMR overlap one another.  Thus, there is literally a chaotic sea
that extends outward from the orbit of Neptune to its 1:2 MMR through
which particles can freely wander.  For example, Fig.~\ref{SD-invade}
shows the results of an experiment where Neptune is on an orbit with
$a=30$~AU and $e=0.2$, and a number of test particles are initially
placed on Neptune crossing orbits ($q<36$~AU, see the top left panel
of Fig.~\ref{SD-invade}). As time passes, objects reach orbits with
larger and larger $q$, and the region cumulatively visited by
particles (painted with small gray dots in the figure) eventually
covers the entire classical Kuiper belt.  As expected, the \Red{1:2
  MMR provides boundary} to this region \Red{(there are objects beyond
  the location of the center of the resonance because the resonance is
  very wide in this simulation)}.  However, no particles are
permanently trapped during this calculation because their trajectories
are time reversible.

The dynamics illustrated in Fig.~\ref{SD-invade} allow us to envision
a mechanism for the implantation of the cold population in the Kuiper
belt in the framework of the Nice model. In fact, as described above,
in the Nice model Neptune undergoes a transient phase during which its
eccentricity is large.  In many of our simulations of this model, this
large eccentricity phase is achieved when Neptune has a semi-major
axis of 27--29~AU, after its last encounter with Uranus. In these
cases, a large portion of what is now the Kuiper belt is already
interior to the location of the 1:2 MMR with Neptune. Thus, it is
unstable, and can be invaded by objects coming from within the
primordial disk (i.e$.$ interior to $\sim 34\,$AU).  When the
eccentricity of Neptune damps out, which takes between $\sim\!0.3$ and
$\sim4\,$Myr, the mechanism that causes this chaos disappears.  The
Kuiper belt becomes stable, and the objects that happen to occupy it
at that time remain trapped for eternity.

\begin{figure}[t!]
\centerline{\psfig{figure=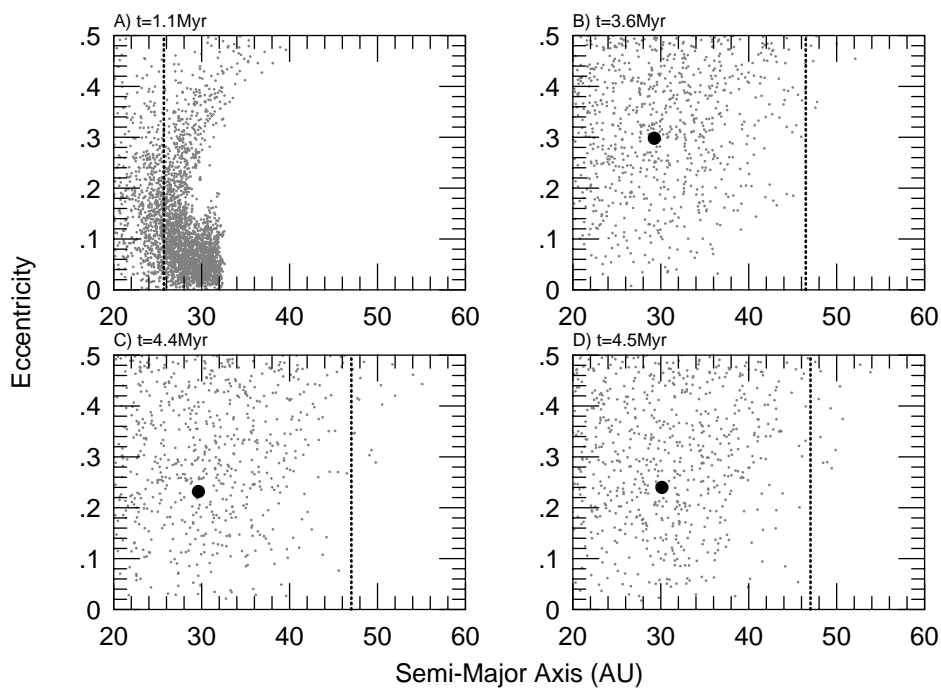,height=14.cm}}
\vspace*{-.3cm}
\caption{Snapshots taken from one of the simulations of the Nice model in
  Tsiganis et al$.$~(2005). The big black dot represents Neptune, the
  small gray dots the particles and the vertical line the location of
  the 1:2 MMR with the planet at the corresponding time (reported in
  the upper left corner of each panel).  For illustrative purposes in
  the last 3 snapshots we have rescaled the semi-major axis unit so
  that the final position of Neptune in this simulation is the current
  one. Time, $t$, is measured since the onset of the instability of
  Uranus and Neptune.} \vspace{0.3cm}
\label{ex-menios}
\end{figure}

Going back to the original simulations of the Nice model, we found
several cases in which the above mechanism for the implantation of
objects into the Kuiper belt was, indeed, at work.
Figure~\ref{ex-menios} gives one example.  Figure~\ref{ex-menios}A
corresponds to the time when the planetary instability has just
started. As one sees, the planetesimal's disk is naturally divided in
two parts: $1)$ the region inside the 1:2 MMR, which is excited as a
result of the previous mean motion resonance sweeping and the onset of
the planetary instability; and $2)$ the region beyond the 1:2 MMR,
which remains relatively dynamically cold.  Figure~\ref{ex-menios}B
shows the system just after the last encounter between Uranus and
Neptune.  Neptune has been kicked outward and onto a large
eccentricity orbit.  Consequently, the 1:2 MMR with Neptune is in the
Kuiper belt.  The main Kuiper belt region ($q>36$~AU and $a<48$~AU) is
totally empty.  Very rapidly, however, particles start to penetrate
into the Kuiper belt due to the reasons described above
(Figure~\ref{ex-menios}C).  When the invasion of particles is
complete, the 1:2 MMR appears as a clear outer boundary of the low
eccentricity population (Figure~\ref{ex-menios}D).

Unfortunately, in all the simulations that we performed of the Nice
model, the number of objects left in the classical Kuiper belt at the
end of the simulations was too small to be analyzed.  We
believe that there are two reasons for this; both related to the fact
that we used a relatively small number of disk particles.  First, we
do not expect the capture efficiency to be very large and thus we
should not have captured many objects.  Perhaps more importantly, the
Neptune's orbital evolution had a stochastic component that was too
prominent, due to the fact that the disk was represented by
unphysically massive objects.  Therefore, in the next section we
perform new simulations where these problems are rectified.

\section{Simulated orbital distributions}
\label{dist}

As described above, self-consistent N--body simulations of the Nice
model, such as those in Tsiganis et al$.$~(2005), simply cannot be
used to study the origin of the Kuiper belt.  Thus, we decided to run
{\it customized} simulations. In these simulations, the planets do not
gravitationally react to the small bodies, which are treated as test
particles.  The desired evolution of the planets' semi-major axes,
eccentricities and inclinations is obtained by the application of
suitable fictitious forces added to their equations of motion.  In
particular, we employ the forces described in Malhotra~(1995) for
evolving semi-major axes and those in Kominami et al$.$~(2005) for
controlling eccentricity and inclination. In addition, since Uranus
and Neptune are often started on crossing orbits and we want to avoid
chaotic evolution due to close encounters, we soften the gravitational
forces (${\bf f}_{UN}$) between these two planets so that
\begin{equation}
    {\bf f}_{UN} = \frac{G m_N m_U}{\left({\bf x}_{UN}^2 +
    \epsilon^2\right)^{3/2}} {\bf x}_{UN},
\end{equation}
where ${\bf x}_{UN}$ is the relative position vector of Uranus and
Neptune, $m_N$ and $m_U$ are their masses, and $\epsilon$ is a
constant which we set to $0.8\,$AU.  In this way, the evolution of the
planetary orbits is smooth.

The above technique has another important advantage --- we have
precise control over the evolution of the planets.  In the purely
N-body simulations in Tsiganis et al$.$~(2005), the evolution of the
orbits of the planets is very chaotic.  Changing the initial
conditions by a very small amount can lead to a drastically different
evolution of the system. Our techniques will allow us to change one
aspect of the evolution from one run to the other, and study the
effect of this one change on the resulting Kuiper belt. Moreover, if
we realize that some simulations need more particles in order to get
adequate statistics, we can run it again with additional test
particles, and the evolution of the planets will not change.

In the following, we describe, in detail, the three main simulations
that we performed, and comment also on two other test runs that we did
for completeness.  Each simulation originally contained 60,000
particles and was run for a billion years. In all, each took at total
of 2 CPU years to complete.  Thus, we could only perform a few runs
and a complete exploration of parameter space is not feasible.

\subsection{Run A}   
\label{runa}

\begin{figure}[t!]
\centerline{\psfig{figure=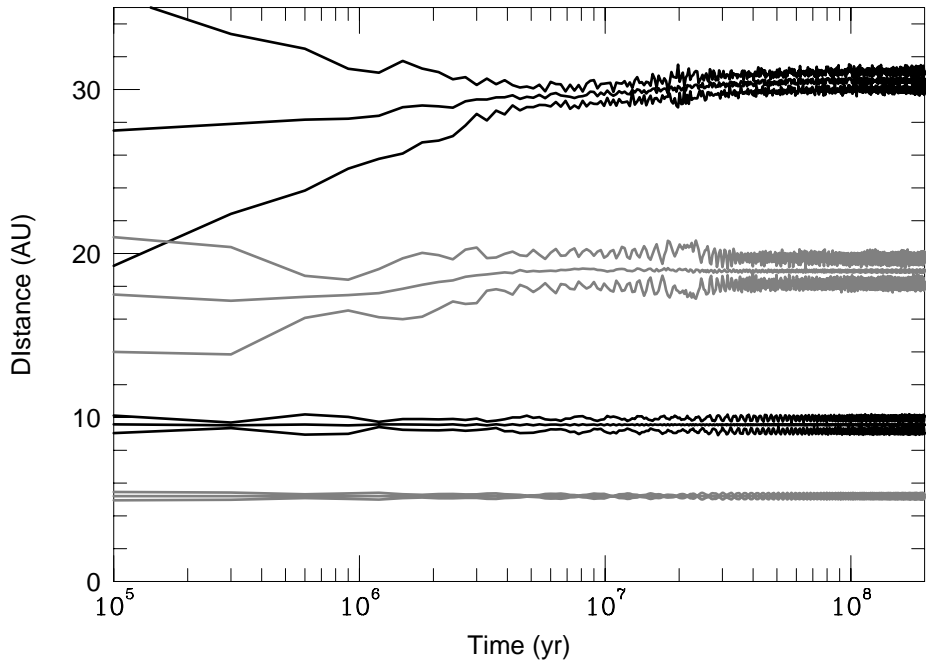,height=14.cm}}
\vspace*{-.3cm}
\caption{The imposed evolution of the giant planets in run A. Each
  planet is represented by three curves. The middle one corresponds to
  the value of the semi-major axis, the lower one to the value of the
  perihelion distance $q$, and the upper one to the value of the
  aphelion distance $Q$.} \vspace{0.3cm}
\label{run1qQ}
\end{figure}

The initial conditions of all our runs are intended to mimic the state
of the system immediately after the last encounter of Neptune with
Uranus.  In Run A, we assume that Neptune has $a=27.5$~AU and $e=0.3$,
and Uranus has $a=17.5$~AU and $e=0.2$. The inclinations of both
planets are small, of order 1 degree, consistent with many of the Nice
model simulations after the last encounter between Uranus and Neptune
(the conservation of the angular momentum tends to decrease the
inclinations of the planets when they scatter each other onto distant
orbits).  Jupiter and Saturn were placed on low-eccentricity,
low-inclination orbits with semi-major axes of 5.2 and $9.6\,$AU,
respectively.  The eccentricities of Uranus and Neptune are assumed to
damp on a timescale of roughly 1~My, although the evolution of the
planets are more complicated than this due, in part, to the
interaction of the planets.  The detailed evolution of the planets is
shown in Fig.~\ref{run1qQ}.

Before we continue, we must discuss one decision we made concerning
the evolution of Uranus' orbit that affects our resulting Kuiper
belts.  In the real Solar System, the $\nu_8$ secular resonance is at
$42\,$AU.  This resonance is very strong and thus can remove objects
on very short timescales (Duncan et al$.$~1995).  Since the location
of this resonance is very sensitive to the semi-major axis of Uranus,
and Uranus is moving in our simulations, we were concerned that the
$\nu_8$ might inadvertently sweep through the classical Kuiper belt
during our integrations thereby unintentionally contaminating the
experiment.  To avoid this, we forced Uranus to migrate a little too
far from the Sun and arrive at its final location a little sooner than
Neptune (see Fig.~\ref{run1qQ}).  Thus, in all our simulations Uranus
final semi-major axis is slightly larger than observed and thus the
$\nu_8$ will be too close to the Sun.

The initial conditions for the disk were inspired by what we see in
the simulations of the Nice model.  As we have seen in
Fig.~\ref{ex-menios}A, the disk is naturally divided into an inner hot
part and an outer cold part, with eccentricities up to 0.2--0.3 and
0.1--0.15, respectively. The boundary between these two disks is at
the location of the 1:2 MMR with Neptune at the time of the planetary
instability, typically at $a\sim 29$~AU.  The inclinations are also
more excited in the inner part than in the outer part.  At the time
when Uranus and Neptune become unstable, the outer part essentially
preserves the initial inclinations of the disk (0.5 degrees in the
simulations of the Nice model).

The initial orbital distribution of our disk particles is an idealized
version of what we see in Fig.~\ref{ex-menios}A. In particular, we
placed particles on orbits with semi-major axes between 20 and 34 AU.
In the inner disk, which we take to be inside of $29\,$AU, the
eccentricity of the particles is assumed to be equal to 0.2, and the
inclination to have a differential distribution of the form
\begin{equation}
\sin(i)~e^{-i^2/2\sigma_i^2},
\label{eq:sig}
\end{equation}
with $\sigma_i=6^\circ$.  Our specific choice of the inclination
distribution in the inner disk is inspired by the results of the Nice
model simulations (Tsiganis et al$.$~2005, see Fig.~\ref{ex-menios}A),
but as we discuss in more detail below, this value is not crucial. In
the outer disk, the eccentricity of the particles is assumed to be
equal to 0.15 and the inclination is assumed to be 0.  An equal number
of particles (30,000) is used to model both the inner and the outer
disk.

Our calculations were performed in two steps.  During the first
time-span of 200~My, the simulations were done using a variant of
swift\_rmvs3 (Levison \& Duncan~1994) that incorporates the fictitious
forces described above.  Thus, the planets migrate and circularize in
these calculations. We then continued the simulation to a billion
years without any migration or damping imposed on the planets, in
order to eliminate scattered disk particles and the bodies that were
marginally unstable in the Kuiper belt.  For this second simulation we
used the code swift\_whm, which executes the algorithm presented in
Wisdom \& Holman~(1991). During the second phase, Particles were
removed from the simulation when they encountered a planet within a
distance of one Hill radius.

\begin{figure}[t!]
\centerline{\psfig{figure=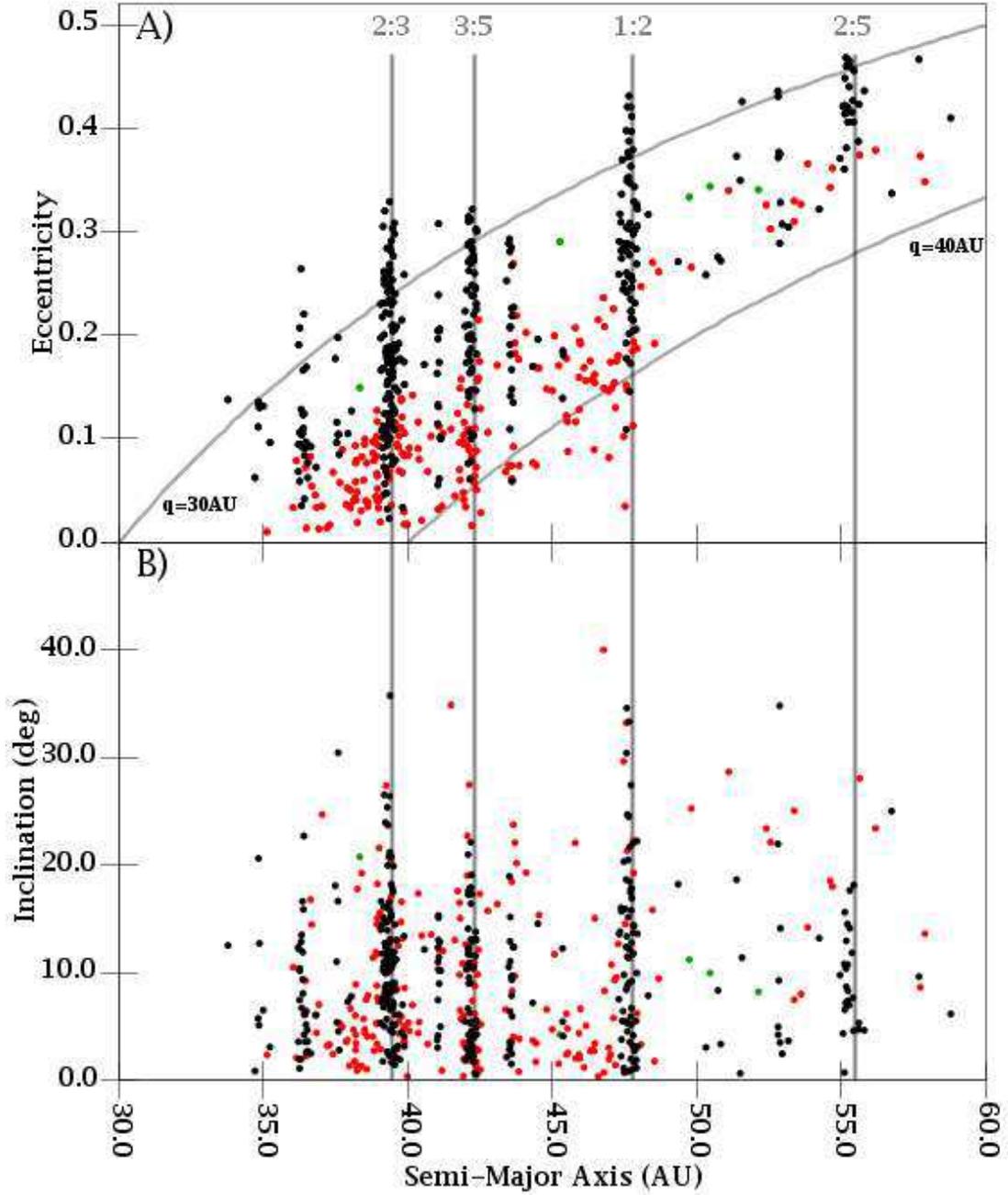,height=17.5cm}}
\vspace*{-.3cm}
\caption{The semi-major axis versus eccentricity (A) and
  semi-major axis versus inclinations (B) distributions of the
  particles captured in the trans-Neptunian region in Run~A.  See
  Fig.~\ref{obs} for a description.}  \vspace{0.3cm}
\label{aei1}
\end{figure}

Fig.~\ref{aei1} shows the $a$--$e$ and $a$--$i$ distributions of the
particles surviving at the end of one billion years.  The color of a
particle indicates its dynamical class.  This is determined by
following the procedures outlined in Gladman et al$.$~(2007), where
each object is classified by a process of elimination using an
integration of a 10 Myr. First, we checked if an object is
`Scattering,' by which we mean that its barycentric semi-major axis
varies by more than $1.5\,$AU over the length of the integration.
Next we look for resonance occupation.  Determining if an object is in
a mean motion resonance with Neptune is not a trivial task.  An object
is classified using the resonant argument, $\phi_{mn}$, for each
$n$:$m$ MMR.  Resonances with $n\!\le\!m\!<\!30$ are considered.  In
addition, a given resonance is checked only if the object in question
is within $0.5\,AU$ of its location.  We split the 10~Myr integration
in to 20 equal sections.  If $\phi$ librates in all of these, by which
we mean it never has $\phi_{mn}$ more than $179.5^\circ$ away from the
libration point, the object is classified as `Resonant'.  If an
object's $\phi_{mn}$ librates in at least 15 of the 20 sections then
it is flagged as possibly resonant and needing to be double checked by
eye. The failure to librate in a given section is often because the
eccentricity of the object drops and $\phi_{mn}$ becomes ill defined.
Fortunately, this only occurs to a small fraction of our objects. In
the figure, the dots that are black represent objects in one of
Neptune's mean motion resonances, dots that are green are `Scattering'
objects, and the red dots are non-resonant stable objects.

We start our discussion with the $a$--$e$ distribution, which is shown
in Fig.~\ref{aei1}A.  This distribution looks similar to the one we
observe (compare with Fig.~\ref{obs}A).  In particular, three main
features are reproduced: (i) the edge of the classical belt at the 1:2
MMR, (ii) the deficiency of nearly-circular objects in the region
immediately interior to the 1:2 MMR, (iii) the extended scattered
disk.  Also, all the main MMR with Neptune are populated, including
those beyond the 1:2.  To our knowledge, this is the first simulation
that reproduces the observed $a$--$e$ distribution of the
trans-Neptunian population so nicely.

Nevertheless, a more accurate comparison between simulation and
observations shows a couple discrepancies:
\begin{itemize}
  
  
\item[$\bullet$] the 40--42 AU region is not depleted, particularly at
  low inclination. The depletion of the real distribution in this
  region is due to the presence of the secular resonances $\nu_8$ and
  $\nu_{18}$ (Duncan et al$.$~1995; Morbidelli et al$.$~1995).  As
  described above, by design these resonances are not present at the
  same location in our model because the orbits of the planets,
  particularly Uranus, are not located exactly at the correct places.
 
\item[$\bullet$] the classical belt (red dots in the figures) appears
  to be somewhat too excited in eccentricity. In particular, although
  the basic shape of the $a$--$e$ distribution follows the
  observations --- nearly-circular objects out to some semi-major axis
  and then $e$ increasing with $a$ --- the location where the
  nearly-circular orbits stop (which we call $a_C$) is too close to
  the Sun.  In particular, real low-eccentricity KBOs can be seen to
  roughly $45\,$AU (Fig.~\ref{obs}), while in Run~A this population
  stops at $43\,$AU.

\end{itemize}

\noindent The last problem is probably the most serious.  In
order to perform a quantitative comparison between the model and the
observed eccentricity distributions in the classical belt, we have
implemented a bias calculator algorithm inspired by the work of
Trujillo \& Brown~(2001). This method takes as input: 1) orbital
element distribution from a model, 2) an assumed absolute magnitude
distribution, and 3) the observed magnitudes and latitudes of the
population being modeled. From this, it predicts what the observed
orbital element distribution of the population should be according to
the model.  It is described in some detail in Morbidelli \&
Brown~(2004), Morbidelli et al.~(2004) and Fernandez \&
Morbidelli~(2006).  We apply it to the population of non-resonant
objects with $42<a<48$~AU in order to avoid the issue, discussed
above, of where we placed the $\nu_8$ secular resonance.  For the real
Kuiper belt objects, we only include objects that have been observed
over multiple oppositions in order to guarantee that the orbits are
well determined.  In the calculation, we assume that each particle
represents full population of objects, with cumulative absolute
magnitude distribution
\begin{equation}
N(<\!H)=C 10^{0.6 H}\ ,
\end{equation}
(Gladman et al$.$~2001), where $C$ is an arbitrary constant (the same
for all particles). We find that if our model were correct, the median
eccentricity of the observed classical belt bodies would be $0.11$,
whereas the median eccentricity of the real objects is $0.07$.
Although not large, this discrepancy is statistically significant.

We can get some insight into how we might fix the above problem by
discussing the origin of $a$--$e$ distribution in our models.  Recall
that this distribution is characterized by low-eccentricity objects
out to $a_C$ and beyond this distance $e$ increases with $a$.
However, $a_C=43\,$AU in Run~A, which is too small.

Also recall that in this simulation, Neptune starts with
$a\!=\!27.5\,$AU and $e\!=\!0.3$. It slowly migrates outward and its
eccentricity decreases with time.  The fact that the eccentricity
damps implies that the region interior to Neptune's 1:2$\,$MMR is most
chaotic at the beginning of the simulation when the resonance is at
$43.6\,$AU.  Thus, the region inside of this distance is filled with
objects.  As the system evolves, the resonance moves outward.  At the
same time, however, Neptune's eccentricity decreases, and so, the
region immediately interior to the 1:2$\,$MMR becomes more stable.  As
a result, we expect fewer low-eccentricity objects beyond $43.6\,$AU
because as the resonance migrates through this region, the chaotic
region shrinks.  This argument predicts that $a_C\!\sim\!43.6\,$AU,
which is roughly what we see.  If it is correct, we can move $a_C$
outward by starting Neptune further from the Sun.  We present such a
simulation in sect.~\ref{ssec:RunB}, below.

The magnitude of the Plutino population is an other important
constraint on the models.  Unfortunately, here the observations are
not clear.  Trujillo et al$.$~(2001) estimate that only roughly 10\%
of the total population of the Kuiper belt are in all the resonances
combined.  On the other extreme, Kavelaars et al$.$~(2007) put the
fraction of objects in Neptune's 2:3~MMR alone at $\sim\!20\%$.  For
this model, we find that 21\% of the particles within $a=50\,$AU are
in Neptune's 2:3~MMR.  Thus, if Kavelaars et al$.$ is correct, our
model is fine.  We address this issue again in sect.~\ref{2to3}.

\begin{figure}[t!]
\centerline{\psfig{figure=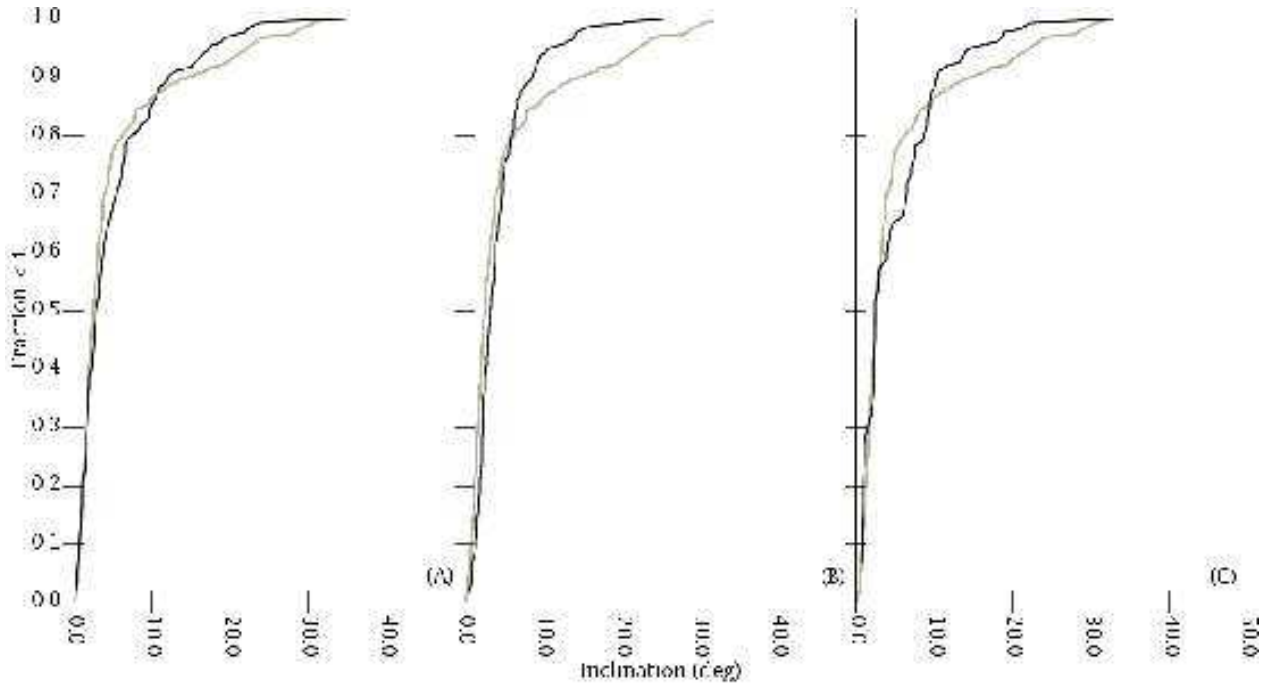,height=9.cm}}
\vspace*{-.3cm}
\caption{The observed cumulative inclination distribution of
  the observed classical belt objects (solid curve) and that expected
  from the result of our simulation (gray curve). Both datasets
  include observational biases. A) run~A. B) run~B. C) run~C.}
\vspace{0.3cm}
\label{i-dist}
\end{figure}

As discussed above, the inclination distribution of the classical belt
is also an important diagnostic.  To compare the distribution obtained
by our model with the observations, we again use the
Trujillo-Brown-like bias calculator mentioned above. As before, we
select real objects and simulation particles that are non-resonant and
have $42<a<48$~AU. In addition, in order to remove the effects of
changes in size distribution with inclination, we only considered
objects with absolute magnitudes, $H$, fainter than 6 (we took a
slightly fainter magnitude than Levison \& Stern~2001's nominal value
in order to be conservative).  The result of this calculation is
presented in Fig.~\ref{i-dist}A.

The black curve in Fig.~\ref{i-dist}A shows the cumulative inclination
distribution of the observed objects.  The gray curve represents the
cumulative inclination distribution expected from our model once the
biases are taken into account using the bias calculator described
above.  As one sees, the agreement is quite good, particularly up to
10 degrees.  This means that we have correctly reproduced the
existence of a cold population --- in particular the inclination
distribution within the cold belt.  We also produce the correct number
of high-inclination objects.  Indeed, the highest inclination object
in our model is $34^\circ$, which can be compared to $31^\circ$ in the
observed dataset.  A Kolmogorov-Smirnov (KS) statistical test (Press
et al$.$~1992) between the two curves in Fig.~\ref{i-dist}A says that
they have a 57\% chance of being derived from the same
distribution\footnote{To be more precise, a KS probability of $0.57$
  means the following.  Assume that there was a single parent
  distribution for both the model and the observations.  In
  particular, the model was a random sample containing, say, $J$
  entries, while the observations contained $K$ entries.  If we were
  to generate two random representations of this parent population,
  one with $J$ entries and one with $K$, there would be a 57\% chance
  that the comparison between these random populations would be {\it
    worse} than what we observe in Fig.~\ref{i-dist}A.  This, despite
  the fact that these new distributions were directly derived from the
  parent.  Therefore, the agreement between our model and the
  observations is very good.  Indeed, any comparison with a KS
  probability greater than $\sim\!0.1$ should be considered
  acceptable, and values as small as $\sim\!0.05$ ($2\sigma$) cannot
  be ruled out.}, which is excellent.

\begin{figure}[t!]
\centerline{\psfig{figure=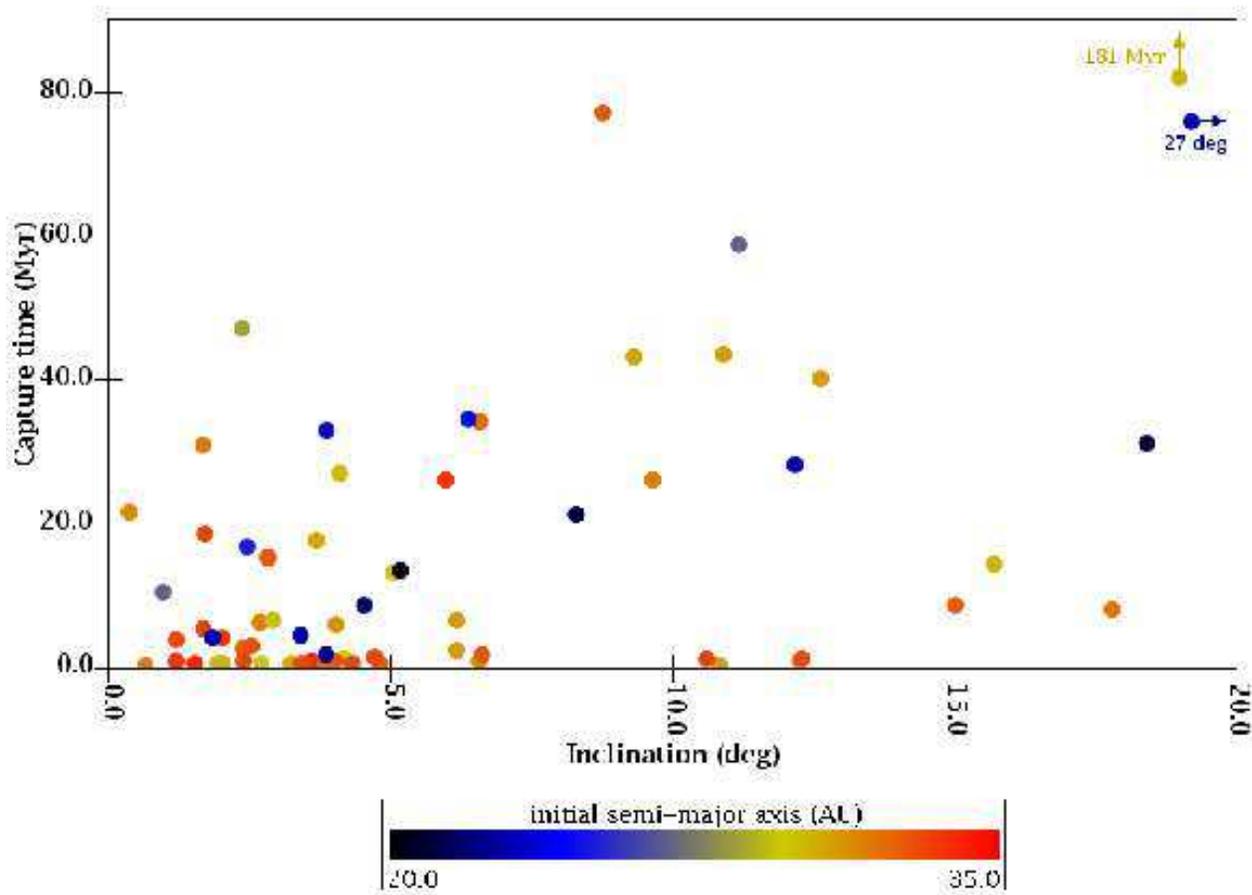,height=12.cm}}
\vspace*{-.3cm}
\caption{For the particles permanently trapped in the classical belt, this
  diagram shows the time of capture as a function of the final
  inclination. The color codes the original semi-major axis of
  the bodies.}  
\vspace{0.3cm}
\label{t-capture}
\end{figure}

Insight into how the low-inclination objects are preserved can be
gleamed by plotting the time at which an object is first captured into
the classical Kuiper belt as a function of its final inclination.  In
particular, Fig.~\ref{t-capture} shows that the implantation of
particles in the Kuiper belt happens in two stages.  The first stage
occurs very rapidly, within the first $\sim\!2\,$My. The particles
that are trapped in the Kuiper belt during this stage do so under the
mechanism described in the previous section. As Fig.~\ref{t-capture}
shows, most of the particles captured at this early time have a low
inclination.  This is due to the fact that they have fewer encounters
with Neptune.  Indeed, we find a direct correlation between the number
of encounters a particle has with Neptune and its final inclination in
the Kuiper belt. There is a well defined cluster with $i<5^\circ$ in
Fig.~\ref{t-capture}, although some particles have inclinations up to
$13^\circ$.

After a few million years the eccentricity of Neptune is damped enough
that the resonances no longer overlap and the bulk of the Kuiper belt
becomes stable. Thus, the capture mechanism described above can no
longer function. The second stage of the capture process starts at
this time, where particles are trapped via the evolutionary pathway
discovered in Gomes~(2003).  As Fig.~\ref{t-capture} shows, the
capture times of the particles during this second stage are more
uniformly distributed, from a few million years up to $200\,$My.  The
inclination distribution of the bodies captured during the second
stage does not show any preference for low inclination.

Fig.~\ref{t-capture} also shows that, while the particles captured
during the first stage come almost exclusively from the outer part of
the disk, those captured during the second stage are mainly from the
inner disk. We will come back to this in sect.~\ref{correlations}.

\subsection{Run B}    
\label{ssec:RunB}

This next run is designed to improve the $a$--$e$ distribution of the
classical belt objects over what was obtained in run~A. In particular,
we aim to extend the distribution of the circular objects trapped in
the classical belt up to $a_C\!\sim\!44\,$AU in order to obtain a
better agreement with the observations.  The only major difference
between this run and run~A was that we placed Neptune initially at
$28.9\,$AU instead of $27.5\,$AU.  We did this so that the initial
location of the 1:2MMR would be slightly beyond $45\,$AU, with the
expectations that the first stage of the capture process --- which
fills the full $a$--$e$ plane up to the resonance's location --- would
implant objects onto nearly-circular orbits up to approximately
44--45~AU.  We also adjusted the migration rate of Neptune so that its
final semi-major axis is at $30.1\,$AU.

\begin{figure}[t!]
\centerline{\psfig{figure=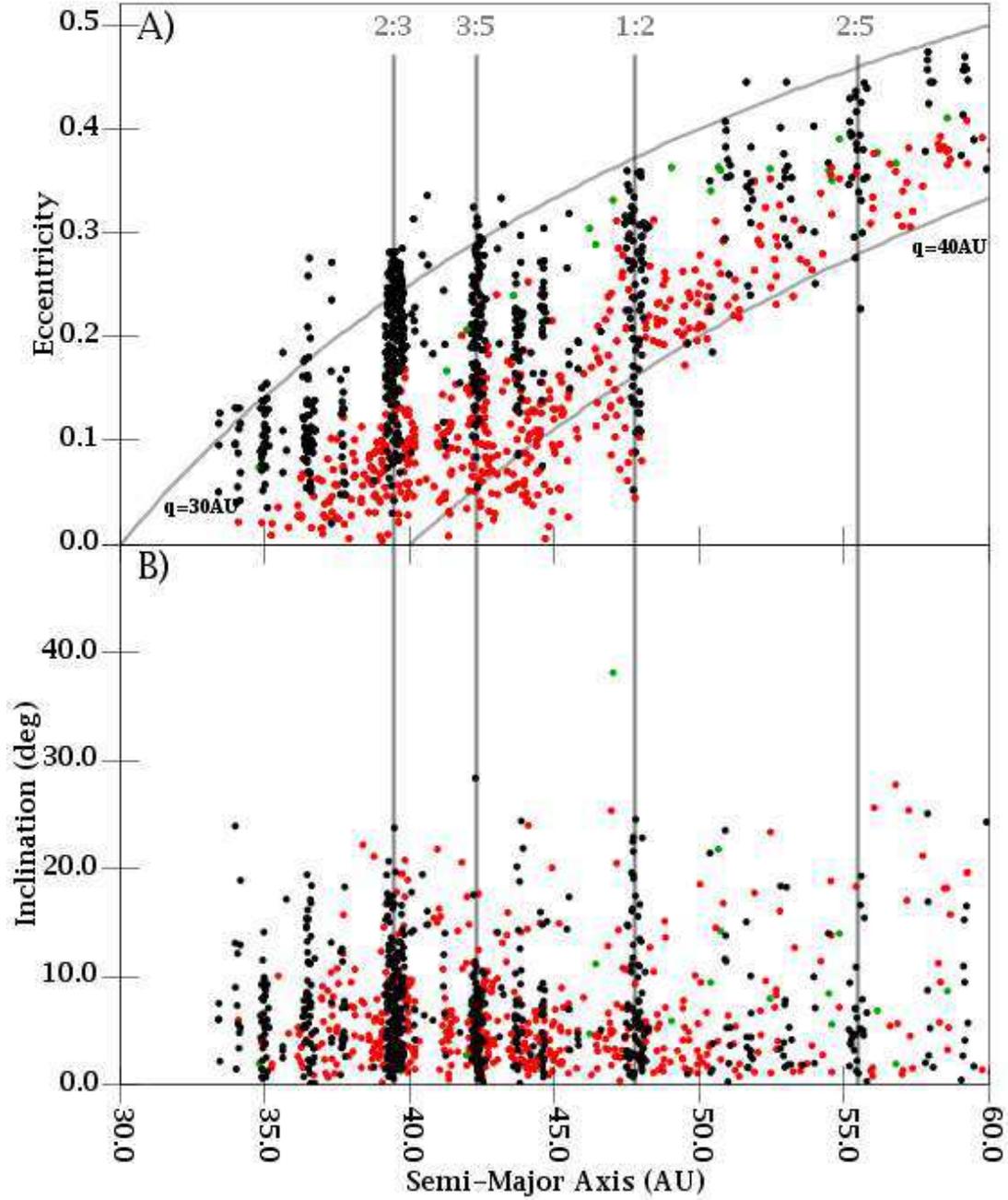,height=17.5cm}}
\vspace*{-.3cm}
\caption{Same as Fig.~\ref{aei1}, but for Run~B.}  \vspace{0.3cm}
\label{aei2}
\end{figure}

The results of this run are shown in Fig.~\ref{aei2}.  Indeed, our
goal seems to have been achieved. The $a$--$e$ distribution is more
similar to the observations than that in run~A (compare with
Fig.~\ref{obs}A).  The population of nearly-circular objects at
low-inclination now extends to $\sim\!45\,$AU.  The deficit of
low-eccentricity objects between 45 and $48\,$AU is preserved, and the
outer edge of the classical belt is still at the final location of the
1:2~MMR with Neptune.  Moreover, beyond the 1:2 resonance, the
extended scattered disk contains objects with larger perihelion
distances than it does in run~A, which is also in better agreement
with the observed distribution.

Nevertheless, when we run the simulated distribution of the classical
belt through the bias calculator, and compare the resulting cumulative
eccentricity distribution with the observed one, the match is still
not perfect: This model predicts that the observed median eccentricity
of the classical population should be 0.10, lower than what run~A
predicts (0.11), but still larger than the observed value of 0.07.
Again, this difference is statistically significant.

Fig.~\ref{i-dist}B shows a comparison between the observed inclination
distribution in the classical belt and the prediction of Run~B after
processing through the bias calculator.  This case is not as good as
that of run~A. The deficit of large inclination bodies appears more
prominent. The observed and the model distributions now diverge at
$i\sim 6^\circ$, and the largest inclination among the captured bodies
is only $25^\circ$. This is probably the consequence of the fact that
we started Neptune at a larger semi-major axis, and thus it migrates a
shorter distance.  This, in turn, partially inhibits Gomes's mechanism
for capturing the hot population.  The KS-test between the observed
and the model distributions tells us that they only have a 4\% chance
of being two statistical representations of the same underlying parent
distribution.  Thus, this model can be ruled out at the $2\sigma$
level based on its inclination distribution. Nevertheless, the
distribution of the inclinations in the cold population is in
excellent agreement with the observations.

At this point, the natural question to ask is whether the poor
inclination distribution of this model is the result of our initial
conditions. Recall that we set $\sigma_i=6^\circ$ (see
Eq.~\ref{eq:sig}) for the inner disk based on the $N$-body simulations
in Tsiganis et al$.$~(2005).  This value is much smaller than the hot
classical belt, which has $\sigma_i\sim 12^\circ$ (Brown~2001).  Thus,
we performed a simulation where we set the initial $\sigma_i$ of the
inner disk to $12^\circ$ in order to see if we produced a more
reasonable inclination distribution.  Surprisingly, the classical belt
inclination distribution of this new run is very similar to the old.
Thus, the poor performance of this model cannot be blamed on our
initial conditions.

\subsection{Run C}    

Another potential way of improving the $a$--$e$ distribution over what
we see in run A, is to increase the damping timescale of Neptune's
eccentricity.  This might help because if Neptune's eccentricity
decays more slowly, the planet, and its 1:2 MMR, have the time to
migrate further out before that the first stage of the trapping
process (the one governed by the planet's eccentricity) ends.
Therefore, in this subsection we present the results of a simulation,
run~C, in which Neptune has the same initial semi-major axis and
eccentricity as in run~A, but the eccentricity damping timescale that
is roughly 3 times longer.

\begin{figure}[t!]
\centerline{\psfig{figure=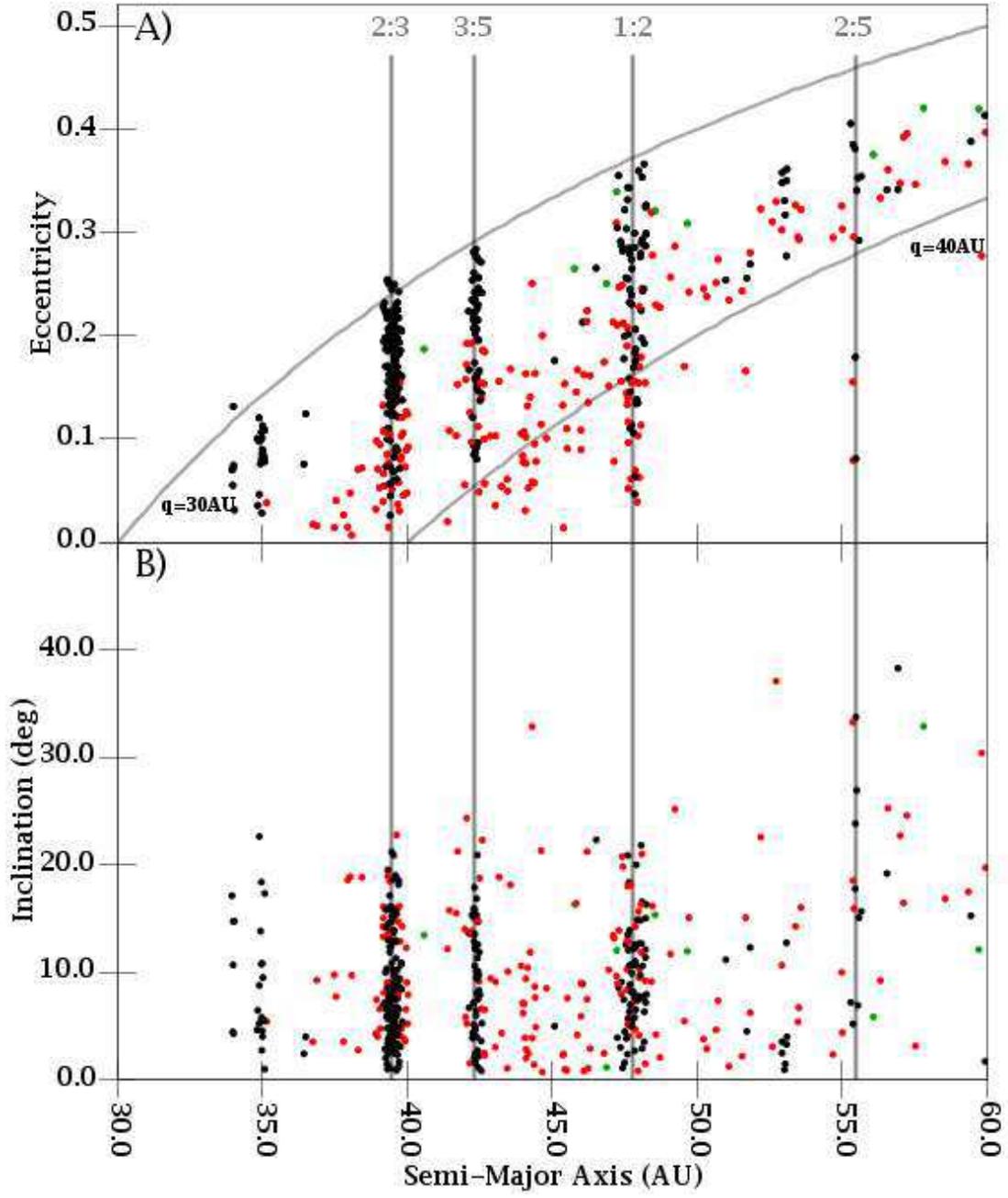,height=17.5cm}}
\vspace*{-.3cm}
\caption{Same as Fig.~\ref{aei1}, but for Run~C.}  \vspace{0.3cm}
\label{aei3}
\end{figure}

The resulting $a$--$e$ distribution is shown in Fig.~\ref{aei3}. Like,
run~B, this distribution matches the observations better than does
run~A in that we have objects with $e\!\sim\!0$ out to $45\,$AU.  The
median eccentricity of the classical objects, according to this model
after accounting for observational biases, should be $\sim 0.13$,
which makes it the worst of the three models.

On the other hand, the inclination distribution appears to be
reasonable.  As Fig.~\ref{i-dist}C shows, the model distribution
appears hotter than the observed distribution between 3 and 10
degrees. This means that there are too many objects at moderate
inclinations (4 to 10 degrees) compared to high inclinations (greater
than 20 degrees).  However, this model does contain objects with
inclinations as large as $33^\circ$ and the KS-test between the two
curves in Fig.~\ref{i-dist}C concludes that the probability that the
two distributions are statistically equivalent is nevertheless 0.35\%.

\subsection{Other runs}

In addition to the three runs described above, we have performed two
additional runs in order to better explore the parameter space and the
effects of these parameters on the resulting Kuiper belt structure.
These runs are not discussed in much detail because they were less
successful than the runs described above.

In run~D, we used the same configuration as in run~A, but we {\it
  decreased} the eccentricity damping timescale by about $1/3$ to
$\sim\!400,000\,$yrs. In other words, run~D stands on the opposite
side of run~A than does run~C with regard to the damping time.  Run~D
did not create a classical Kuiper belt at all.  This result is not
surprising, because as Fig.~\ref{SD-invade} shows, it takes roughly a
million years for the particles to penetrate the Kuiper belt, and
thus, Neptune's eccentricity damps too fast in this run so that the
particles do not have enough time to penetrate into the Kuiper belt
before that the latter becomes stable.
 
Run~E was set up to enhance the effect of Gomes~(2003) mechanism.  In
particular, we set Neptune's initial semi-major axis to $28\,$AU. In
addition, Neptune's eccentricity damping rate was set to $3\,$Myrs
(like run~C).  However, Neptune's migration was done in two stages.
For the first $5\,$Myr, Neptune migrated as it did in run~B, so that
at the end of this stage it was at $29.2\,$AU.  Then, Neptune migrated
very slowly, reaching $30.1\,$AU in about $100\,$Myr.  In this run we
found that the population of nearly circular classical bodies extends
to larger semi-major axes, almost up to the 1:2 MMR location. The
extended scattered disk beyond the 1:2 MMR extends to larger
perihelion distance, so that there is almost no apparent `edge' at the
1:2 MMR in the resulting $a$--$e$ distribution.  Furthermore, the
inclination distribution is too excited, so that the cold population
is under-represented.  Nevertheless, the number of bodies captured at
high inclination is not significantly higher than it was in run~B.
Thus, the goal of this simulation was not achieved.

\section{Origin of the correlations between physical and dynamical properties}
\label{correlations}

As we reviewed in the introduction, the Kuiper belt displays two
strong correlations between an object's physical properties and its
orbit. The first one is a relationship between absolute magnitude (or
size) and inclination. In particular, the existence of a cold `core'
in the inclination distribution is visible only for those bodies with
$H\!\gtrsim\!6$ (Levison \& Stern, 2001). Most bodies with $H\!>\!6$
have $i\!<\!4^\circ$. The second correlation is between colors
and inclination.  The cold population is deficient of objects with a
neutral spectral gradient, often called `gray objects' (Tegler \&
Romanishin~2000; Trujillo \& Brown~2002; Doressoundiram et
al$.$~2005).  Because colors can be affected by evolutionary processes
such as irradiation, heating, out-gassing, collisional resurfacing, we
think that the correlation between size and inclination has better
chances to be directly related to the origin and primordial sculpting
of the Kuiper belt.  Therefore, we first discuss the $H$ versus $i$
correlation.

\begin{figure}[t!]
\centerline{\psfig{figure=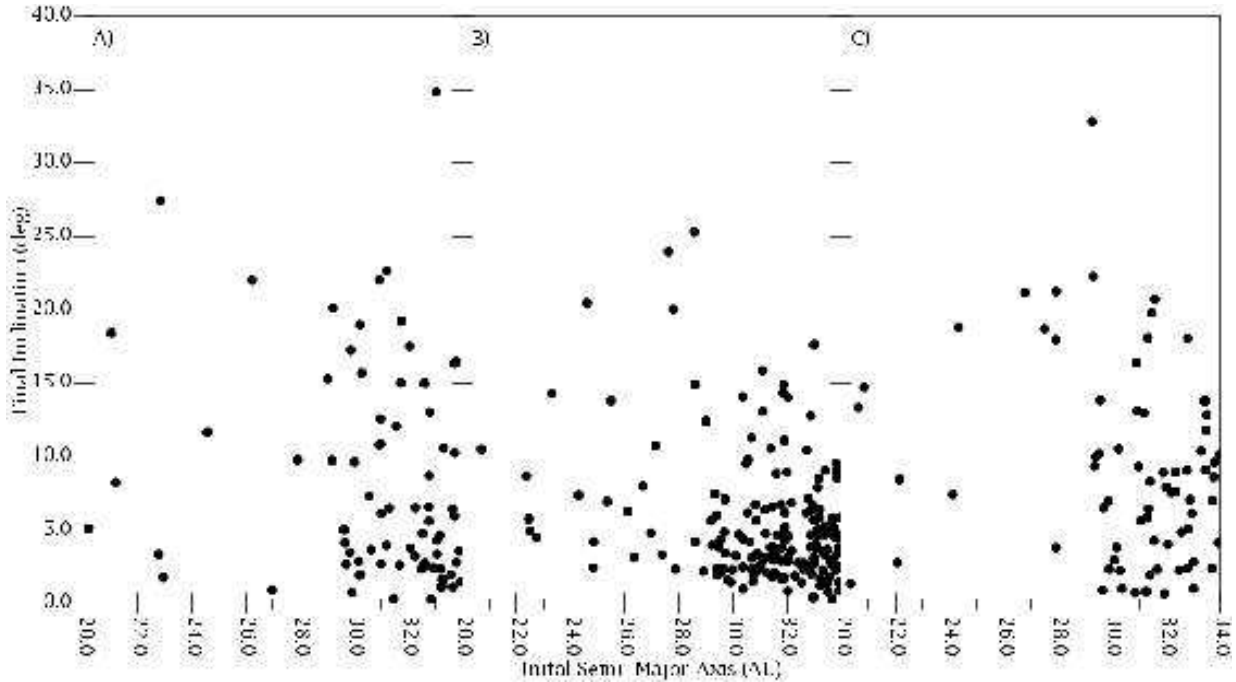,height=9.cm}}
\vspace*{-.3cm}
\caption{The final inclination as a function of the initial semi-major axis,
  for particles trapped in the classical Kuiper belt. A) from run~A.
  B) from run~B. C) from Run~C.}  \vspace{0.3cm}
\label{a0i}
\end{figure}

Fig.~\ref{a0i} shows the final inclination of the particles trapped in
the classical belt as a function of their initial semi-major axis
($a_0$), for our three main runs. In both Runs~B and C, the low
inclination `core' (namely an over-density of objects with
$i<4^\circ$) is made of particles with $a{_0}\!>\!29$~AU.  In our most
extreme case (Run~B), only 5\% of the particles with $i\!<\!4^\circ$
initially come from the region interior to $29\,$AU (we call this
fraction $f_{\rm cold}$), while most of the objects with
$i\!>\!15^\circ$ come from this region ($f_{\rm hot}=0.54$).  For
Run~A these fractions are $f_{\rm cold}=0.07$ and $f_{\rm hot}=0.17$,
and they are $f_{\rm cold}=0.11$ and $f_{\rm hot}=0.40$ for Run~C.
So, although there is large variation from run to run, there is always
a trend that the low indication particles formed further from the Sun
then the high inclination ones.

Before we discuss the implications of the above result, we would like
to understand its origin.  It is partially a consequence of our
initial conditions, which are based on the N-body simulations of the
Nice model, where the outer disk is initially cold (low eccentricities
and inclinations), whereas the inner disk is already excited.
However, this cannot be the sole explanation.  If it were, we would
expect that captured large-inclination Kuiper belt objects would have
started the simulation with large inclinations in the inner disk.
This is not seen.  Therefore, the deficiency of low inclination
particles from the inner disk cannot be simply explained by our
initial inclination distribution.

Another important effect is that objects from the inner disk encounter
Neptune more often than those from the outer disk.  For example in
run~B, the captured classical Kuiper belt objects that originated in
the inner disk encountered Neptune an average of 59 times, while those
from the outer disk only encounter Neptune 19 times on average.
Therefore, the orbital excitation due to the close encounters with
Neptune is much more pronounced for objects from the inner disk, and
so the chances to preserve an initial low inclination are smaller.

What ever its exact origin, the above result can explain the
correlation between $H$ versus $i$.  It is natural to expect that the
size distribution of the disk planetesimals was a function of
heliocentric distance. In particular, given that the timescale for the
growth of a body increases with increasing orbital period
(Safronov~1969), it is legitimate to expect that some bodies in the
inner part of the disk could acquire larger sizes than the largest of
the bodies in the outer part of the disk.  Let us assume, for example,
that bodies with $H<6$ could form only on orbits with, say, $a<28$~AU.
Fig.~\ref{a0i} shows that, for these bodies, the inclination
distribution would be rather extended, with no low inclination `core'
--- at least for Runs~B and C.  Therefore, our simulations explain, at
least qualitatively, the $H$ versus $i$ correlation.

\begin{figure}[t!]
\centerline{\psfig{figure=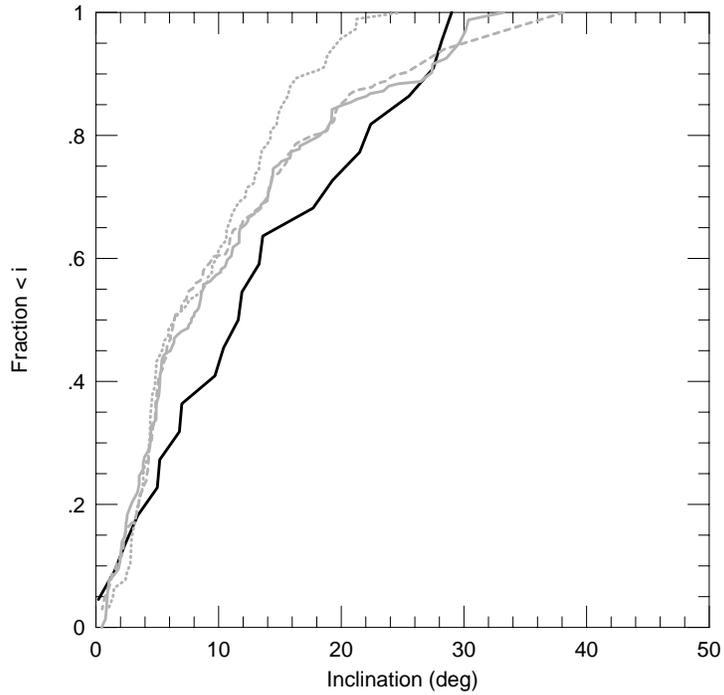,height=15.cm}}
\vspace*{-.3cm}
\caption{The cumulative observed inclination distribution for the
  bright objects in the Kuiper belt. The black curve shows all
  multi-opposition objects with $32<a<50\,$AU, $q>31\,$AU, and
  $H<5.5$.  The dashed, solid, and dotted gray curves show fictitious
  objects from Runs~A, B, and C, respectively, which were trapped in
  the same region of space.  For the models, however, we only included
  those objects with initial $a<27\,$AU.  In addition, we ran the
  models through our survey simulator.}  \vspace{0.3cm}
\label{ihot}
\end{figure}

From a more quantitative point of view, the black curve in
Fig.~\ref{ihot} shows the inclination distribution of all
multi-opposition Kuiper belt objects with $H<5.5$ (to be conservative,
we took a slightly brighter magnitude than Levison \& Stern's limit).
In order to get better statistics, here we included all objects with
$32<a<50\,$AU and $q>31\,$AU.  For comparison, the gray curves plot
what we would expect from our three main runs assuming that
$H\!<\!5.5$ objects only formed interior to $27\,$AU.  The KS
probabilities are 0.42, 0.24, and 0.17 for runs A, B, and C,
respectively, and thus the agreement is very good.  Thus, our
model supports the idea that big objects, or at least intrinsically
bright ones, originated closer to the Sun.

We now come to the issue of the colors. Let us assume, for simplicity,
that there are no evolutionary processes, and, for some unknown
reason, the objects that formed in the inner parts of the disk are
gray and those from the outer parts are red.  As we discussed above,
our simulations produce a classical Kuiper belt where a higher
fraction of objects with $i<4^\circ$ come from outer disk than the
high inclination objects.  This fact is most pronounced in run~B,
where $f_{\rm cold}=0.05$ and $f_{\rm hot}=0.54$.  Thus, this model
predicts that 95\% of the cold population would be red, while the hot
population would be roughly 50-50.

Nevertheless, even run~B would predict that we should be finding some
gray objects among the cold population, and our other runs would argue
for significantly more.  Observations, conversely, seem to show a
total absence of gray objects in the cold `core'.  Having said this,
we believe that the total lack of gray low-inclination objects is
either a statistical fluke, or, more likely, an indication that the
colors are not primordial.  This is due to the fact that, independent
of the Nice model, it is difficult to imagine a dynamical mechanism
that can create a population of objects {\it only} at high
inclination.  Any mechanism that we can think of will naturally
include some low-inclination objects and therefore there should be
some gray bodies in the cold population if color is primordial.

As described above, there is also a correlation between color and
perihelion distance in that all objects with $q\!>\!39$~AU are red
(Doressoundiram et al$.$~2005).  The bodies coming from the inner disk
tend to acquire final perihelion distances that are smaller than the
objects from the outer disk.  Therefore, if we look at bodies with
$q>39$~AU, the ratio of outer disk to inner disk bodies should be
smaller than, say, for $36<q<39$.  In fact, we find that in run~B all
the bodies trapped in the Kuiper belt with $q>40$~AU come from the
outer disk.  Thus, if colors are primordial and are a function of
formation heliocentric distance, our models can explain these
observations.

In conclusion, the results are somewhat mixed when it comes to whether
our models can produce the relationship between orbit and physical
characteristics.  As Fig$.$~\ref{ihot} shows, in all three models the
inclination distribution of objects that formed interior to $27\,$AU
is consistent with the observed distribution of intrinsically bright
KBOs.  However, only run~B, and perhaps run~C, can convincingly
explain the correlation between colors and inclination.  As we
discussed above, we believe it likely that some evolutionary
processes, yet to be understood, created, or at the very least,
sharpened the correlations between color and orbit.  If true, these
processes must be something like collisional resurfacing, that has
been active since the observed Kuiper belt was put in place.

\section{Additional Considerations}
\label{discussion}

In this section we compare our model to other important features of
the trans-Neptunian population.

\subsection{The final mass of the Kuiper belt}
\label{mass}

In run~A, 180 particles out of 60,000 remained permanently captured in
\Red{non-resonant orbits} with $a\!<\!48\,$AU.  The Nice model
requires that the primordial planetesimal disk was
$\sim\!35\,M_\oplus$ and that roughly $\sim\!25\,M_\oplus$ remained in
the disk at the time of the instability.  Combining these numbers we
find that run~A predicts that the mass of the classical belt should be
$\sim\!0.07\,M_\oplus$. In run~B, we find that 360 particles are
trapped, and 125 in run~C.  Thus, our model predicts that the
classical Kuiper belt should contain between $\sim\!0.05$ and
$\sim\!0.14 M_\oplus$.  These results are of the same order as the
observed mass of the Kuiper belt, which is estimated to be in the
range 0.01--0.1~$M_\oplus$ (Gladman et al$.$~2001; Bernstein et
al$.$~2004), \Red{of which between 80\% and 90\%} is in the classical
population (Trujillo et al$.$~2001; \Red{Kavelaars et al$.$~2007}).
Thus, our model explains the mass deficit of the Kuiper belt.

Of course, to be viable, our model needs to explain not only the total
mass of the belt, but also the total number of bright, detectable
bodies found there. This supplies an important constraint on the
original size distribution in the planetesimal disk.  In fact it is
obvious that, if the original size distribution was such that the bulk
of the mass was carried by small bodies (meters to few kilometers in
size), the captured population would not have a sufficient number of
detectable objects.

Given that the trapping efficiency in the classical belt is of order
$10^{-3}$, our model requires that there were $\sim\!1000$ Pluto-size
bodies in the original planetesimal disk.  This is in agreement with
the expectations (for example see Stern~1991) that are based on the
low probabilities of: 1) the capture of Triton by Neptune (for example
see Agnor \& Hamilton~2006), 2) the formation of the Pluto--Charon
binary by an energetic collision of two large objects (Canup~2005),
and 3) the existence of Pluto-size bodies (e$.$g$.$ Eris) in the
scattered disk\footnote{Independent of how it originally formed, the
  scattered disk represents not more than 1\% of the pristine disk
  population (see for instance Duncan \& Levison~1997).}.  It might be
possible to argue that each of these low probability events, taken
individually, could occur at random even if the original number of
Pluto-size objects were small.  However, we strongly feel that the
fact that all the three events happened almost proves that there were
originally a large number of Plutos.

Because the dynamical evolution of the disk particles is independent
of size, our model requires that the original size distribution in the
disk was similar to that currently observed in the Kuiper belt, namely
one which breaks from a steep to a shallow slope at about $50$--100~km
in diameter (Bernstein et al$.$~2004).  This size distribution fits in
nicely with the Nice model for two reasons.  First, O'Brien et
al$.$~(2005) showed a $35\,M_\oplus$ disk, like the one required by
the Nice model, would not be significantly altered by collisions
during the $\sim\!600\,$Myr that preceded the LHB if it had this
size-distribution. Moreover, Charnoz \& Morbidelli (2006) showed that
such a disk can explain the total number of comet-size bodies in the
Oort cloud and in the scattered disk.

\subsection{Orbital distribution of the Plutinos}
\label{2to3} 

Up to now we have considered only the orbital distribution in the
classical belt. However, another important diagnostic of any model is
whether it can reproduce the orbital element distribution inside the
major mean motion resonances with Neptune, in particular the 2:3.  In
fact, in contrast with all previous scenarios of Kuiper belt
formation, our model does not include mean motion resonance sweeping
of a cold disk of planetesimals.  In all our runs, the initial
location of the 2:3 MMR is beyond the outer edge of the particle disk,
and thus, there is no contribution from the mechanism proposed by
Malhotra~(1993, 1995).  Therefore, the investigation of the
distribution of Plutinos can provide an important venue for comparing
our model to previous scenarios.

\begin{figure}[t!]
\centerline{\psfig{figure=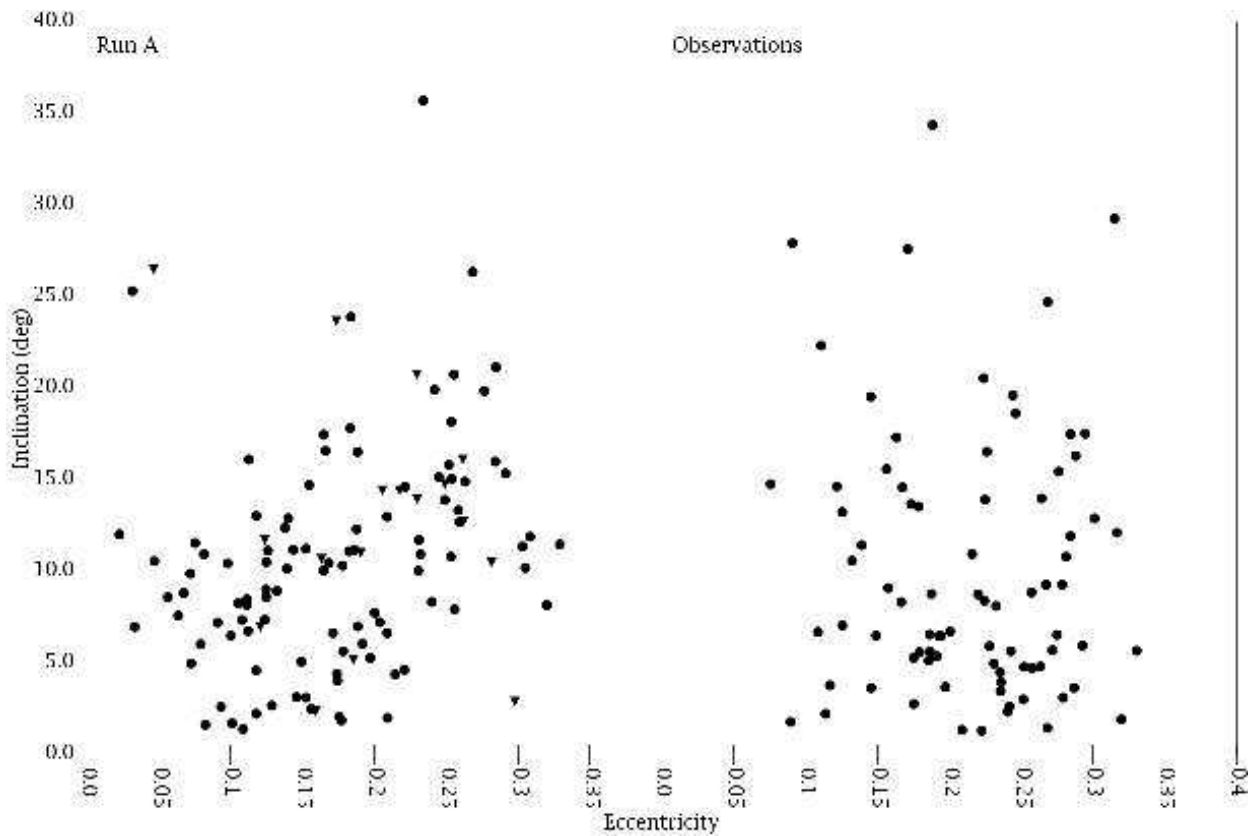,height=11.cm}}
\vspace*{-.3cm}
\caption{The eccentricity--inclination distribution of the Plutinos. Left
  panel: the result of run~A; dots refer to particles from the outer
  disk and triangles to particles from the inner disk. Right panel:
  the observed distribution.}  \vspace{0.3cm}
\label{2:3}
\end{figure}

Fig.~\ref{2:3} compares the $e$--$i$ distribution of the Plutinos
obtained in run~A, against the observed distribution.  When studying
this figure it is important to note that the observed distribution
suffers from observational biases, while the model distribution does
not.  Unfortunately, we cannot account for observational biases as we
did for the classical belt because of the dynamics of the resonance
itself.  Our bias calculation assumes a uniform distribution of all
orbital angles and that there is no correlation between inclination,
eccentricity, and these angles.  For objects in the 2:3 MMR there is a
strong correlation between the argument of perihelion and the values
of the eccentricity and inclination due to a strong Kozai effect
(Kozai~1962; Williams \& Bensen~1971; Morbidelli et al$.$~1995).  As a
result, the observed eccentricity and inclination distributions are
very sensitive to exactly where in the sky the telescopes were pointed
(Gladman, pers. comm.).  This information is not available.

In Fig.~\ref{2:3} there is a good agreement between the two
inclination distributions, at least by eye.  The major difference seen
between the two distributions in the figure is in the eccentricity
distribution --- the eccentricities are larger in the observations.
This could be the result of the observational biases.

In the left panel of Fig.~\ref{2:3}, the dots refer to particles
captured from the outer disk and the triangles to particles from the
inner disk. Outer disk particles dominate the distribution slightly.
The ratio between inner disk and outer disk particles is similar to
the one obtained for the hot population of the classical belt.
Therefore, our model predicts that the Plutinos and the hot population
should share the same physical characteristics --- at least when it
comes to those characteristics related to formation location.  In
addition, the cold population should be different.  More data is
needed before we can do such an analysis.

Despite some difficulties and uncertainties, the distribution of the
Plutinos obtained in our model is much better than in any other
previous model. In the models of Malhotra~(1995) and Hahn \&
Malhotra~(2005), the inclination distribution of the Plutinos should
be similar to that of the pre-migrated planetesimal disk interior to
$40\,$AU.  Although Malhotra~(1995) demonstrated that some of the
bodies captured in the 2:3 resonance can acquire large inclinations
(or an inclination different from its pristine one), Gomes~(2000), who
studied in detail the inclination excitation mechanism, concluded that
these changes would not modify the inclination distribution enough to
explain the observations.  So, in the models by Malhotra~(1995) and
Hahn \& Malhotra~(2005), the only way to explain the fact that the
Plutinos and the classical belt beyond $42\,$AU have different $i$
distributions --- the Plutinos lack the cold core --- is if there was
a sudden break in the inclination distribution of pre-migrated
planetesimal disk at the current location of the 3:2 MMR.  Given that
there is no physical reason why this break should be near the 3:2
MMR's current location\footnote{To be clear, the fact that the
  inclination distribution of the Plutinos is different from the
  classical belt is not discussed by Malhotra~(1995) or Hahn \&
  Malhotra~(2005).  The main conclusion of these works is that the
  resonance capture does not significantly change the inclination
  distribution and so the observed inclination distribution in the
  resonances must represent the state of the disk before the planets
  migrated.  They do not study the excitation of the disk. The rest of
  this discussion is our interpretation of this result.}, this
juxtaposition must be a coincidence according to these models.  This
seems unlikely.  In Gomes~(2003), the Plutinos are a mixture of bodies
trapped from the scattered disk, originally formed closer to Neptune,
and bodies trapped from a more distant cold disk as in
Malhotra~(1995).  Thus, we would expect the same inclination
distribution and the same correlations between physical
characteristics and orbits in the Plutinos as we see in the classical
belt.  This is not observed.

The fact that the Plutinos do not have a low-inclination core and that
the distribution of physical properties of the Plutinos is comparable
to that of the hot population are important constraints for any model.
These characteristics are achieved in our model because of two
essential ingredients: (i) the assumption of a truncated disk at
$\sim\!34$~AU and (ii) the fact that Neptune `jumps' directly to
27--28~AU.  As a result, the 2:3~MMR does not migrate through the
disk, but instead jumps over it.

\subsection{Mean motion resonance populations beyond 50 AU}
\label{dist-res} 

In the final $a$--$e$ distributions obtained in run~A,~B and~C (see
Figs.~\ref{aei1}, \ref{aei2}, and \ref{aei3}), there are clearly
particles trapped in mean motion resonances beyond $50\,$AU.  This
includes objects in the 4:9 ($51.7\,$AU), 3:7 ($53\,$AU), 2:5
($55.4\,$AU), and the 3:8 MMRs ($57.9\,$AU). The 2:5 MMR has the most
prominent population.  This is consistent with the observed
distribution in the belt (Chiang et al$.$~2003).

\begin{figure}[t!]
\centerline{\psfig{figure=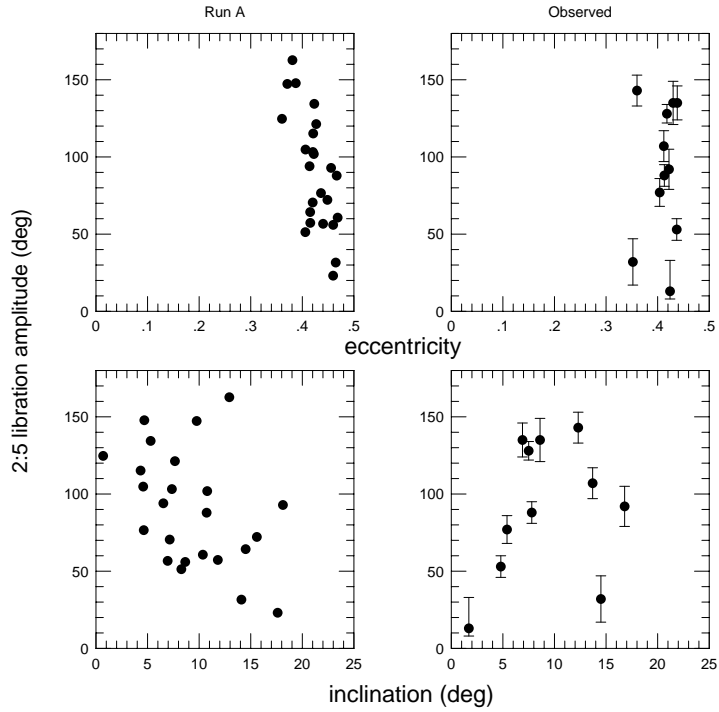,height=15.cm}}
\vspace*{-.3cm}
\caption{Projections of the eccentricity--inclination-- libration
  amplitude distribution of objects in Neptune's \Red{2:5} MMR.  The left
  two panels show the results from run~A, while the right two panels
  show the observations.  The libration amplitudes of the real objects
  is taken from Lykawka \& Mukai~(2006).}  \vspace{0.3cm}
\label{5to2}
\end{figure}

In a recent paper, Lykawka \& Mukai~(2006) studied the orbital
distribution of objects in these resonances and pointed out that these
distributions supply important constraints for any Kuiper belt
formation scenario.  In particular, they argue that the libration
amplitude distribution can be an important diagnostic.  Thus, here we
compare the results of our model to the observations, concentrating,
as suggested by Lykawka \& Mukai, on the 2:5 MMR with Neptune. In
Fig.\ref{5to2}, we show the eccentricity, inclination, and libration
amplitude of objects in this resonance.  The left two panels report
the results from run~A, while the right two panels show the
observations.  The libration amplitudes for the real 2:5 librators is
taken from Lykawka \& Mukai (2006).  Visually, there is excellent
agreement between the model and the observations.  (Unfortunately, for
reason discussed in sect.~\ref{2to3}, we cannot do a quantitative
comparison) We consider this result an important accomplishment of our
model.

\section{Conclusions}
\label{conclusions}

In this paper, we have studied the origin and orbital evolution of the
Kuiper belt in the context of the Nice model for the orbital evolution
of the giant planets.  Recall that the Nice model explains, for the
first time, the orbital architecture of the giant planets (Tsiganis et
al$.$~2005), the origin of the late heavy bombardment in the inner
solar system (Gomes et al$.$~2005), the existence of the Trojans of
Jupiter (Morbidelli et al$.$~2005) and of Neptune (Tsiganis et al$.$
2005; Trujillo \& Sheppard~2006), and the origin of at least some of
the giant planet irregular satellites (Nesvorn\'y et al$.$~2006).

Based on the simulations of the Nice model we \Red{presuppose that the
  proto-planetary disk was truncated at $\sim\!30\,$AU so that Neptune
  does not migrate too far (see Gomes et al$.$~2004).  In addition, we
  assume that} Neptune was scattered outward by Uranus to a semi-major
axis between 27 and 29~AU and an eccentricity of $\sim\!0.3$, after
which its eccentricity damped on a timescale of roughly $1\,$Myr.
\Red{Furthermore,} we assume the inclinations of the planets remained
small during this evolution.  Given these premises, we find that our
simulations reproduce the main observed properties of the
trans-Neptunian population: 1) the co-existence of resonant and
non-resonant populations, 2) the peculiar $a$--$e$ distribution of the
classical belt, 3) the existence of an outer edge at the location of
the 1:2 MMR with Neptune, 4) the bi-modal inclination distribution of
the classical population, 5) the correlations between physical
properties and inclination, 6) the orbital distribution of the
Plutinos and the \Red{2:5} librators, 7) the existence of the extended
scattered disk, and, last but not least, 8) the mass deficit.

%

Our models suffer from a significant problem, however: The
eccentricities are too large in our classical belt.  The median
eccentricity that we obtain is 0.10--0.13, whereas the observed value
is 0.07.  Although this problem is significant, we believe that it
cannot be used as an argument against this scenario. When looking at
the Kuiper belt at the level of detail we have attempted here (which,
in itself, is unprecedented), the set of planetary evolutions that are
consistent with the observed structure becomes extremely narrow and we
probably have not been able to pin-point it yet.  This seems
particularly likely given that we have only been able to perform a few
simulations and held some free parameters constant (like Neptune's
initial eccentricity).  It is also possible that some differences are
just a statistical fluke, or some are due to missing physics in our
simulations (such as stochasticity during planet migration or
collisional damping).  Only future models will tell.

Although when it comes to using the Kuiper belt to unravel the history
of the planets the devil is in the details, we think that the list of
successes of our model outweighs the problems that remain open.  No
other model ever reproduced the observed Kuiper belt nearly as closely
as the one in this study.  Indeed, only Hahn \& Malhotra~(2005), who
studied the classical resonance sweeping scenario, have attempted the
type of comparison that we presented here.  They find that the list of
successes of their model is much more limited. It explains the
existence of resonant bodies and their eccentricity range. But it
fails to explain the $(e,i)$ distribution inside the resonances. It
requires an independent and unspecified excitation mechanism to
explain the $(e,i)$ distribution in the classical belt and the
existence of objects in the resonances beyond 50 AU. It cannot explain
the mass deficit of the classical belt, but it has to rely on yet
another mechanism (collisional grinding) for it. It cannot explain why
the outer edge of the Kuiper belt is so close to the 1:2 MMR of
Neptune so that it has to invoke a coincidence. So, an improbable
patchwork of models is required in order to explain the Kuiper belt as
a whole in the scenario.

Thus, we tentatively conclude that the structure of the Kuiper belt is
an additional, substantial argument in support of the Nice model. If
our conclusion is correct, and the Nice model is valid, then the
Kuiper belt is the relic of the primordial massive planetesimal disk
that surrounded the planets and that triggered a late instability of
the outer planetary system.  This instability caused the lunar late
heavy bombardment. The Kuiper belt acquired its present observed
characteristics during that time, which was a seismic shake-up that
totally reshaped the solar system structure.

\medskip\medskip

\acknowledgments HFL is grateful for funding from NASA's Origins, OPR,
and PGG programs.  AM acknowledges funding from the french National
Programme of Planetaology (PNP).  R$.$G$.$ is grateful to Conselho
Nacional de Desenvolvimento Cient\'ifico e Tecnol\'ogico for the
financial support. We would also like to thank Brett Gladman for
suppling us with the ability to dynamically classify our objects, and
to Bill Bottke and David Nesvorn\'y for useful discussions.

\section*{References}

\begin{itemize}
\setlength{\itemindent}{-30pt}
\setlength{\labelwidth}{0pt}

\item[] Agnor, C.~B., Hamilton, D.~P.\ 2006.\ Neptune's capture of its
  moon Triton in a binary-planet gravitational encounter.\ Nature 441,
  192-194.

\item[] Allen, R.~L., Bernstein, G.~M., Malhotra, R.\ 2001.\ The Edge
  of the Solar System.\ Astrophysical Journal 549, L241-L244.

\item[] Allen, R.~L., Bernstein, G.~M., Malhotra, R.\ 2002.\ 
  Observational Limits on a Distant Cold Kuiper Belt.\ Astronomical
  Journal 124, 2949-2954.
  
\item[] Bernstein, G.~M., Trilling, D.~E., Allen, R.~L., Brown, M.~E.,
  Holman, M., Malhotra, R.\ 2004.\ The Size Distribution of
  Trans-Neptunian Bodies.\ Astronomical Journal 128, 1364-1390.
  
\item[] Brasser, R., Duncan, M.~J., Levison, H.~F.\ 2006.\ Embedded
  star clusters and the formation of the Oort Cloud.\ Icarus 184,
  59-82.

\item[] Brown, M.~E.\ 2001.\ The Inclination Distribution of the
  Kuiper Belt.\ Astronomical Journal 121, 2804-2814.
  
\item[] Canup, R.~M.\ 2005.\ A Giant Impact Origin of Pluto-Charon.\ 
  Science 307, 546-550.
  
\item[] Charnoz, S., Morbidelli, A.\ 2006.\ Coupled Dynamical And
  Collisional Evolution Of The Oort Cloud And The Kuiper Belt.\ 
  AAS/Division for Planetary Sciences Meeting Abstracts 38, \#34.04.
  
\item[] Chiang, E.~I., Jordan, A.~B., Millis, R.~L., Buie, M.~W.,
  Wasserman, L.~H., Elliot, J.~L., Kern, S.~D., Trilling, D.~E.,
  Meech, K.~J., Wagner, R.~M.\ 2003.\ Resonance Occupation in the
  Kuiper Belt: Case Examples of the 5:2 and Trojan Resonances.\ 
  Astronomical Journal 126, 430-443.

\item[] Doressoundiram, A., Barucci, M.~A., Romon, J., Veillet, C.\ 
  2001.\ Multicolor Photometry of Trans-neptunian Objects.\ Icarus
  154, 277-286.
  
\item[] Doressoundiram, A., Barucci, M.~A., Tozzi, G.~P., Poulet, F.,
  Boehnhardt, H., de Bergh, C., Peixinho, N.\ 2005.\ Spectral
  characteristics and modeling of the trans-neptunian object (55565)
  2002 AW$_{197}$ and the Centaurs (55576) 2002 GB$_{10}$ and (83982)
  2002 GO$_{9}$: ESO Large Program on TNOs and Centaurs.\ Planetary
  and Space Science 53, 1501-1509.
 
\item[] Duncan, M.~J., Levison, H.~F., Budd, S.~M.\ 1995.\ The
  Dynamical Structure of the Kuiper Belt.\ Astronomical Journal 110,
  3073-3081.
  
\item[] Duncan, M.~J., Levison, H.~F.\ 1997.\ A scattered comet disk
  and the origin of Jupiter family comets.\ Science 276, 1670-1672.
  
\item[] Elliot, J.~L., and 10 colleagues 2005.\ The Deep Ecliptic
  Survey: A Search for Kuiper Belt Objects and Centaurs. II. Dynamical
  Classification, the Kuiper Belt Plane, and the Core Population.\ 
  Astronomical Journal 129, 1117-1162.
  
\item[] Fern{\'a}ndez, J.~A., Morbidelli, A.\ 2006.\ The population of
  faint Jupiter family comets near the Earth.\ Icarus 185, 211-222.

\item[] Gladman, B., Kavelaars, J.~J., Petit, J.-M., Morbidelli, A.,
  Holman, M.~J., Loredo, T.\ 2001.\ The Structure of the Kuiper Belt:
  Size Distribution and Radial Extent.\ Astronomical Journal 122,
  1051-1066.

\item[] Gladman, B., Marsden, B.~G., VanLaerhoven, C$.$ 2007.\ 
  Nomenclature in the Outer Solar System.\ Kuiper Belt, in press.

\item[] Gomes, R.~S.\ 2003.\ The origin of the Kuiper Belt
  high-inclination population.\ Icarus 161, 404-418.
  
\item[] Gomes, R.~S., Morbidelli, A., Levison, H.~F.\ 2004.\ Planetary
  migration in a planetesimal disk: why did Neptune stop at 30 AU?.\ 
  Icarus 170, 492-507.
  
\item[] Gomes, R., Levison, H.~F., Tsiganis, K., Morbidelli, A.\ 
  2005.\ Origin of the cataclysmic Late Heavy Bombardment period of
  the terrestrial planets.\ Nature 435, 466-469.
  
\item[] Grundy, W.~M., Noll, K.~S., Stephens, D.~C.\ 2005.\ Diverse
  albedos of small trans-neptunian objects.\ Icarus 176, 184-191.

\item[] Hahn, J.~M., Malhotra, R.\ 1999.\ Orbital Evolution of Planets
  Embedded in a Planetesimal Disk.\ Astronomical Journal 117,
  3041-3053.

\item[] Hahn, J.~M., Malhotra, R.\ 2005.\ Neptune's Migration into a
  Stirred-Up Kuiper Belt: A Detailed Comparison of Simulations to
  Observations.\ Astronomical Journal 130, 2392-2414.
  
\item[] Haisch, K.~E., Jr., Lada, E.~A., Lada, C.~J.\ 2001.\ Disk
  Frequencies and Lifetimes in Young Clusters.\ Astrophysical Journal
  553, L153-L156.
  
\item[] Holman, M.~J., Wisdom, J.\ 1993.\ Dynamical stability in the
  outer solar system and the delivery of short period comets.\ 
  Astronomical Journal 105, 1987-1999.
  
\item[] Kavelaars, J.~J., Jones, L., Gladman, B., Parker, J.~W.,
  Petit, J.-M. 2007.\ The orbital and spatial distribution of the
  Kuiper belt.\ Kuiper Belt, in press.

\item[] Kenyon, S.~J., Luu, J.~X.\ 1998.\ Accretion in the Early
  Kuiper Belt. I. Coagulation and Velocity Evolution.\ Astronomical
  Journal 115, 2136-2160.
  
\item[] Kenyon, S.~J., Luu, J.~X.\ 1999a.\ Accretion in the Early
  Kuiper Belt. II. Fragmentation.\ Astronomical Journal 118,
  1101-1119.
  
\item[] Kenyon, S.~J., Luu, J.~X.\ 1999b.\ Accretion in the Early
  Outer Solar System.\ Astrophysical Journal 526, 465-470.
  
\item[] Kenyon, S.~J., Bromley, B.~C.\ 2004a.\ Stellar encounters as
  the origin of distant Solar System objects in highly eccentric
  orbits.\ Nature 432, 598-602.

\item[] Kenyon, S.~J., Bromley, B.~C.\ 2004b.\ The Size Distribution of
  Kuiper Belt Objects.\ Astronomical Journal 128, 1916-1926.
  
\item[] Kominami, J., Tanaka, H., Ida, S.\ 2005.\ Orbital evolution
  and accretion of protoplanets tidally interacting with a gas disk.\ 
  Icarus 178, 540-552.
  
\item[] Kozai, Y.\ 1962.\ Secular perturbations of asteroids with high
  inclination and eccentricity.\ Astronomical Journal 67, 591.
  
\item[] Levison, H.~F., Duncan, M.~J.\ 1994.\ The long-term dynamical
  behavior of short-period comets.\ Icarus 108, 18-36.

\item[] Levison, H.~F., Duncan, M.~J.\ 1997.\ From the Kuiper Belt to
  Jupiter-Family Comets: The Spatial Distribution of Ecliptic Comets.\ 
  Icarus 127, 13-32.
  
\item[] Levison, H.~F., Stern, S.~A.\ 2001.\ On the Size Dependence of
  the Inclination Distribution of the Main Kuiper Belt.\ Astronomical
  Journal 121, 1730-173
  
\item[] Levison, H.~F., Stewart, G.~R.\ 2001.\ Remarks on Modeling the
  Formation of Uranus and Neptune.\ Icarus 153, 224-228.
  
\item[] Levison, H.~F., Morbidelli, A.\ 2003.\ The formation of the
  Kuiper belt by the outward transport of bodies during Neptune's
  migration.\ Nature 426, 419-421.
  
\item[] Lubow, S.~H., Seibert, M., Artymowicz, P.\ 1999.\ Disk
  Accretion onto High-Mass Planets.\ Astrophysical Journal 526,
  1001-1012.
  
\item[] Lykawka, P.~S., Mukai, T.\ 2006.\ Evidence for an excited
  Kuiper belt of 50AU radius in the first Myr of Solar system
  history.\ Icarus, in press.
  
\item[] Malhotra, R.\ 1993.\ The Origin of Pluto's Peculiar Orbit.\ 
  Nature 365, 819.

\item[] Malhotra, R.\ 1995.\ The Origin of Pluto's Orbit: Implications
  for the Solar System Beyond Neptune.\ Astronomical Journal 110, 420.
  
\item[] Masset, F., Snellgrove, M.\ 2001.\ Reversing type II
  migration: resonance trapping of a lighter giant protoplanet.\ 
  Monthly Notices of the Royal Astronomical Society 320, L55-L59.
  
\item[] Morbidelli, A., Thomas, F., Moons, M.\ 1995.\ The resonant
  structure of the Kuiper belt and the dynamics of the first five
  trans-Neptunian objects.\ Icarus 118, 322.

\item[] Morbidelli, A.\ 2002.\ Modern Integrations of Solar System
  Dynamics.\ Annual Review of Earth and Planetary Sciences 30, 89-112.
  
\item[] Morbidelli, A., Emel'yanenko, V.~V., Levison, H.~F.\ 2004.\ 
  Origin and orbital distribution of the trans-Neptunian scattered
  disc.\ Monthly Notices of the Royal Astronomical Society 355,
  935-940.
  
\item[] Morbidelli, A., Brown, M.~E.\ 2004.\ The kuiper belt and the
  primordial evolution of the solar system.\ Comets II 175-191.

\item[] Morbidelli, A., Levison, H.~F.\ 2004.\ Scenarios for the
  Origin of the Orbits of the Trans-Neptunian Objects 2000 CR$_{105}$
  and 2003 VB$_{12}$ (Sedna).\ Astronomical Journal 128, 2564-2576.
  
\item[] Morbidelli, A., Levison, H.~F., Tsiganis, K., Gomes, R.\ 
  2005.\ Chaotic capture of Jupiter's Trojan asteroids in the early
  Solar System.\ Nature 435, 462-465.

\item[] Morbidelli, A., Crida, A.\ 2007.\ The dynamics of Jupiter and
  Saturn in the gaseous proto-planetary disk.\ Icarus, submitted.
  
\item[] Murray-Clay, R.~A., Chiang, E.~I.\ 2006.\ Brownian Motion in
  Planetary Migration.\ Astrophysical Journal 651, 1194-1208.

\Red{
\item[] Nesvorn{\'y}, D., Vokrouhlick{\'y}, D., Morbidelli, A.\ 2007.\ 
  Capture of Irregular Satellites during Planetary Encounters.\ 
  Astronomical Journal 133, 1962-1976.
}
  
\item[] O'Brien, D.~P., Morbidelli, A., Bottke, W.~F.\ 2005.\ 
  Collisional Evolution of the Primordial Trans-Neptunian Disk:
  Implications for Planetary Migration and the Current Size
  Distribution of TNOs.\ Bulletin of the American Astronomical Society
  37, 676.
  
\item[] Pollack, J.~B., Hubickyj, O., Bodenheimer, P., Lissauer,
  J.~J., Podolak, M., Greenzweig, Y.\ 1996.\ Formation of the Giant
  Planets by Concurrent Accretion of Solids and Gas.\ Icarus 124,
  62-85.
  
\item[] Press, W., Teukolsky, S.A., Vetterling, W.T., Flannery, B.P.\ 
  1992.  Numerical Recipes in Fortran.  The art of scientific
  computing, 2nd edition.\ Cambridge University Press.
  
\item[] Safronov, V.S.\ 1969.\ `Evolution of the Protoplanetary Cloud
  and the Formation of the Earth and Planets.\ Nauka Press.

\item[] Sheppard, S.~S., Trujillo, C.~A.\ 2006.\ A Thick Cloud of
  Neptune Trojans and Their Colors.\ Science 313, 511-514.
  
\item[] Stern, S.~A.\ 1991.\ On the number of planets in the outer
  solar system - Evidence of a substantial population of 1000-km
  bodies.\ Icarus 90, 271-281.

\item[] Stern, S.~A.\ 1996.\ On the Collisional Environment, Accretion
  Time Scales, and Architecture of the Massive, Primordial Kuiper
  Belt..\ Astronomical Journal 112, 1203.
  
\item[] Stern, S.~A., Colwell, J.~E.\ 1997a.\ Accretion in the
  Edgeworth-Kuiper Belt: Forming 100-1000 KM Radius Bodies at 30 AU
  and Beyond..\ Astronomical Journal 114, 841.

\item[] Stern, S.~A., Colwell, J.~E.\ 1997b.\ Collisional Erosion in
  the Primordial Edgeworth-Kuiper Belt and the Generation of the 30-50
  AU Kuiper Gap.\ Astrophysical Journal 490, 879.
  
\item[] Tegler, S.~C., Romanishin, W.\ 2000.\ Extremely red
  Kuiper-belt objects in near-circular orbits beyond 40 AU.\ Nature
  407, 979-981.

\Red{
\item[]Tera F, Papanastassiou, D.A., Wasserburg, G.J.\ 1974.\ Isotopic
  Evidence for a terminal lunar cataclysm.\ Earth Planet Sci.\ Lett.\ 
  22, 1-21.
}

\item[] Thommes, E.~W., Duncan, M.~J., Levison, H.~F.\ 2003.\ 
  Oligarchic growth of giant planets.\ Icarus 161, 431-455.

\item[] Trujillo, C.~A., Brown, M.~E.\ 2001.\ The Radial Distribution
  of the Kuiper Belt.\ Astrophysical Journal 554, L95-L98.
  
\item[] Trujillo, C.~A., Jewitt, D.~C., Luu, J.~X.\ 2001.\ Properties
  of the Trans-Neptunian Belt: Statistics from the
  Canada-France-Hawaii Telescope Survey.\ Astronomical Journal 122,
  457-473.
  
\item[] Trujillo, C.~A., Brown, M.~E.\ 2002.\ A Correlation between
  Inclination and Color in the Classical Kuiper Belt.\ Astrophysical
  Journal 566, L125-L128.
  
\item[] Tsiganis, K., Gomes, R., Morbidelli, A., Levison, H.~F.\ 
  2005.\ Origin of the orbital architecture of the giant planets of
  the Solar System.\ Nature 435, 459-461.

\item[] Wisdom, J., Holman, M.\ 1991.\ Symplectic maps for the n-body
  problem.\ Astronomical Journal 102, 1528-1538.

\end{itemize}

\clearpage

\end{document}